\documentclass[journal,twoside]{asmb}

\begin{document}
\title{RIS-aided Mixed RF-FSO Wireless Networks: Secrecy Performance Analysis with Simultaneous Eavesdropping}

\author{
\IEEEauthorblockN{Md. Mijanur Rahman, A. S. M. Badrudduza, \textit{Member, IEEE}, Noor Ahmad Sarker, Md. Ibrahim, \textit{Graduate Student Member, IEEE}, and Imran Shafique Ansari, \textit{Senior Member, IEEE}}
}

\twocolumn[
\begin{@twocolumnfalse}
\maketitle
\begin{abstract}
\section*{Abstract}

The appearance of sixth-generation networks has resulted in the proposal of several solutions to tackle signal loss. One of these solutions is the utilization of reconfigurable intelligent surfaces (RIS), which can reflect or refract signals as required. This integration offers significant potential to improve the coverage area from the sender to the receiver. In this paper, we present a comprehensive framework for analyzing the secrecy performance of a RIS-aided mixed radio frequency (RF)-free space optics (FSO) system, for the first time. Our study assumes that a secure message is transmitted from a RF transmitter to a FSO receiver through an intermediate relay. The RF link experiences Rician fading while the FSO link experiences Málaga distributed turbulence with pointing errors. We examine three scenarios: 1) RF-link eavesdropping, 2) FSO-link eavesdropping, and 3) a simultaneous eavesdropping attack on both RF and FSO links. We evaluate the secrecy performance using analytical expressions to compute secrecy metrics such as the average secrecy capacity, secrecy outage probability, strictly positive secrecy capacity, effective secrecy throughput, and intercept probability. Our results are confirmed via Monte-Carlo simulations and demonstrate that fading parameters, atmospheric turbulence conditions, pointing errors, and detection techniques play a crucial role in enhancing secrecy performance.
\end{abstract}

\begin{IEEEkeywords}
\section*{Keywords} 

Rician fading, Málaga fading, physical layer security, reconfigurable intelligent surface, pointing error

\end{IEEEkeywords}
\end{@twocolumnfalse}
]

\section{Introduction}
\subsection{Background and Literature Study}

As the sixth generation ($6$G) of wireless communication approaches, the possibility of utilizing reconfigurable intelligent surfaces (RIS) to address the negative impacts of wireless channels is being explored as a crucial technology \cite{di2020smart}. However, to create a truly intelligent environment, there is a significant strategy under consideration, which is to have control over the wireless medium \cite{9134962}. To address this need, a RIS has been developed using passive components that can be programmed and managed through a RIS controller allowing it to reflect signals toward specific directions as required \cite{do2021multi}. Furthermore, the mixed radio frequency (RF)-free space optical (FSO) systems are considered potential structures for the next-generation wireless networks \cite{soleimani2015generalized}. The use of RIS in both RF and FSO transmissions can help solve signal blockage issues that arise in wireless communication.

FSO communications are seen as a promising option that can provide fast data transfer speeds and be applied in a range of scenarios such as serving as a backup to fiber, supporting wireless networks for back-haul, and aiding in disaster recovery efforts \cite{10058969}. However, they are vulnerable to pointing errors and atmospheric conditions and are not suitable for transmitting information over long distances. Through the implementation of relaying strategy, the dual-hop RF-FSO mixed models merge the strengths of RF and FSO communication technologies \cite{lei2020secure, sarker2021intercept}. In \cite{zedini2016performance}, the authors demonstrated how pointing errors, atmospheric turbulence, and path loss affect a mixed FSO-RF system and provided insights for improving the design and operation of such systems. The authors of \cite{yang2017unified} derived analytical expressions for the outage probability (OP), average data rate, and ergodic capacity (EC) of the RF-FSO system, and assessed its performance in the presence of multiple users with varying data rate requirements. Recently, the authors of \cite{qu2022uav,ashrafzadeh2019framework,6952039,6777774} enhanced the dual-hop performance by optimizing the system parameters and making it suitable for space-air-ground integrated networks.

There has been a lot of research in the literature that studied single RIS-aided systems \cite{zhang2021performance,selimis2021performance,9609960,al2021ris,ruku2023effects,xu2021performance,tasci2022new,yang2021performance}. In \cite{zhang2021performance}, the accuracy and effectiveness of RIS-assisted systems in modifying wireless signals were evaluated by assuming practical factors such as phase shift and amplitude response that can affect their performance. Researchers at \cite{selimis2021performance} demonstrated that RIS could improve system performance over a Nakagami-$m$ fading channel by examining signal-to-noise ratio (SNR) and channel capacity. The findings of the study also provide insights into how to optimize RIS-empowered communications in practical scenarios. The system performance of a RIS-aided network is assessed by the authors in \cite{9609960} wherein they suggest that the number of reflecting elements used in the network does not affect the diversity gain. On the other hand, the system performance of RIS-aided dual-hop network is analyzed in \cite{9424709,9057633,9347448,9124704, malik2022performance}. For example, the authors of \cite{9424709} conducted a study comparing RIS-equipped RF sources and RIS-aided RF sources, and suggested that mixed RF-FSO relay networks utilizing these two types of sources offer great potential for enhancing the performance of wireless communication networks in various environments, both indoors and outdoors. In \cite{9057633}, the authors concluded that incorporating RIS in mixed FSO-RF systems can greatly enhance the coverage area. This is achieved by improving the signal quality and reducing the signal attenuation that may occur during transmission. However, it has been observed that in dual-hop communication systems with co-channel interference, RIS can help mitigate the impact of interference from nearby channels \cite{9347448}. Ref. \cite{9124704} proposed a study on the effect of different system parameters, including the number and placement of RIS elements, on the performance of a RIS-assisted communication system. Here, the authors concluded that the most effective RIS configuration for optimal performance depends on the specific communication scenario and network requirements.

Wireless communications face a significant challenge in terms of protecting the privacy of information because their inherent characteristics make them vulnerable to security threats \cite{badrudduza2021security}. Till date, the security of wireless communication has relied on different encryption and decryption techniques that take place in the higher levels of the protocol stack \cite{zhang2017securing}. Newly suggested physical layer security (PLS) methods are now seen as a practical solution to stop unauthorized eavesdropping in wireless networks by utilizing the unpredictable nature of time-varying wireless channels \cite{9231044}. Recently, extensive research has been conducted to explore the secrecy performance of mixed RF-FSO systems. The authors of \cite{erdogan2021secrecy} concluded that using a mixed model offers better security compared to using RF or FSO technology alone, and they emphasized the importance of implementing appropriate security measures and techniques. Another study in \cite{islam2020secrecy} examined the secrecy performance of a mixed RF-FSO relay channel with variable gain while \cite{sarker2021effects} provided insights into the secrecy performance of a cooperative relaying system, and emphasized the significance of selecting suitable statistical models and security techniques. Additionally, researchers in \cite{wang2022secrecy} evaluated the performance of the mixed RF-FSO system with a wireless-powered friendly jammer and analyzed the impact of different system parameters on secrecy performance. However, some challenges and limitations associated with the dual-hop model were identified in \cite{saber2021secrecy} including the impact of atmospheric turbulence on the FSO link's performance and the importance of accurate channel estimation. Finally, a new model for the mixed RF-FSO channel was proposed in \cite{islam2021impact} that takes into account arbitrary correlation, and the results showed that both correlation and pointing error could significantly affect the secure outage performance of the model. The potential of RIS to improve confidentiality in wireless networks has not been extensively studied in the context of RIS-assisted RF-FSO systems. Nonetheless, studies such as \cite{zhuang2022secrecy} analyzed the secrecy performance of a non-orthogonal multiple access-based FSO-RF system while investigating the impact of imperfect channel state information. Both studies provide valuable insights into enhancing the secrecy performance of wireless communication systems.


\subsection{Motivation and Contributions}
Although RIS-aided mixed RF-FSO systems are strong contenders for upcoming $6$G wireless networks and diverse applications, there has been limited investigation into their capacity to maintain secrecy in the available literature. The current literature mainly focuses on mixed RF-FSO systems and does not fully investigate the security performance of RIS-assisted RF-FSO systems particularly when RIS is used in both links. In this paper, the authors conduct a PLS analysis of the RIS-aided RF-FSO system configuration and evaluate its secrecy performance under the simultaneous influence of RF and FSO eavesdropping attacks, which, to the best of the authors' knowledge, has not been inspected before for this type of configuration. In addition, since wireless channels experience frequent variation over time, assuming a Rician channel in the RF links would provide a more realistic environment to model the wireless propagation perfectly \cite{salhab2021accurate}. Meanwhile, the Málaga fading distribution applied to the FSO link in the system being examined produces reliable results, particularly in challenging atmospheric turbulence and pointing error scenarios \cite{islam2021impact}. Motivated by these advantages, we introduce a secure scenario over the Rician-Málaga mixed RF-FSO fading model. The key contribution of this research is given below.

\begin{itemize}
\item In the past few decades, numerous studies have investigated the secrecy performance of mixed RF-FSO systems, such as \cite{erdogan2021secrecy,islam2020secrecy,sarker2021effects,wang2022secrecy,saber2021secrecy,islam2021impact,ghosh2023secrecy,pan2019secrecy}. However, the secrecy performance of dual-hop systems that incorporate both RF and FSO links aided by RIS remains an open concept, with no research conducted on this specific configuration to date. In this paper, we propose a RIS-aided mixed RF-FSO network in the presence of two different eavesdroppers accounting for their ability to intercept information transmitted through both RF and FSO links.

\item Firstly, we obtain the cumulative distribution function (CDF) of the dual-hop RF-FSO system under decode-and-forward (DF) relaying protocols by utilizing the CDF of each link. Furthermore, we develop the new analytical expressions of average secrecy capacity (ASC), lower bound of secrecy outage probability (SOP), probability of strictly positive secrecy capacity (SPSC), effective secrecy throughput (EST), and intercept probability (IP). These expressions are novel compared to the previous works as the proposed model is completely different from the existing RF-FSO literature.

\item The expressions that we derived have been utilized to generate numerical results with specific figures. Furthermore, we have confirmed the precision of the analytical results through Monte-Carlo (MC) simulations. This validation through simulation strengthens the reliability of our analysis.

\item In an effort to increase the practicality of our analysis, we have provided insightful remarks that shed light on the design of secure RIS-aided mixed RF-FSO relay networks. To ensure a more realistic analysis, we have taken into account the major impairments and features of both RF and FSO links. For instance, we have incorporated the impacts of fading parameters and the number of reflecting elements for RF links, as well as atmospheric turbulence, detection techniques, and pointing error conditions for FSO links.
\end{itemize}

\subsection{Organization}
The paper is organized into several sections. Section II provides an introduction to the models of the system and channel that are utilized in the study. In Section III, the paper presents analytical expressions for five significant performance metrics including ASC, SOP, and the probabilities of SPSC, EST, and IP. Section IV is particularly interesting as it features enlightening discussions and numerous numerical examples. Finally, Section V serves as the conclusion to the paper.

\section{System Model and Problem Formulation}
 \begin{figure*}[!ht]
\centerline{\includegraphics[width=0.6\textwidth,angle =0]{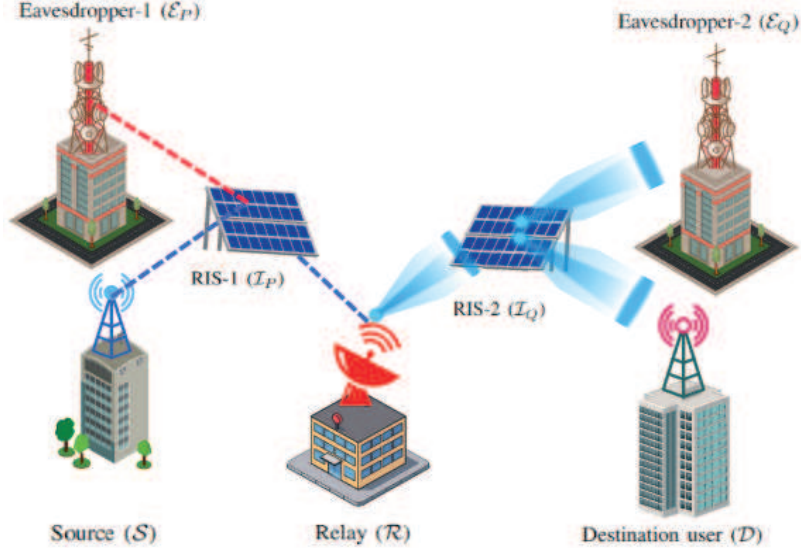}}
\vspace{0mm }
\caption{System model of a combined RIS-aided dual-hop RF-FSO system with source ($\mathcal{S}$), relay ($\mathcal{R}$), the destination user ($\mathcal{D}$), and eavesdropper ($\mathcal{E}$ ).}
\label{pm1}
\end{figure*}
As depicted in Fig. \ref{pm1}, we present the system model of a RIS-aided combined RF-FSO DF-based relaying system where RIS-RF system forms the first hop and the second hop is composed of RIS-FSO system. Since it is unlikely that the $\mathcal{S}$ and $\mathcal{R}$ could communicate directly due to obstructions, the $\mathcal{S}$ links to the $\mathcal{R}$ through a RIS ($\mathcal{I}_{P}$) mounted on a structure. Similarly, communication between $\mathcal{R}$ and $\mathcal{D}$ is established through another RIS ($\mathcal{I}_{Q}$). RIS works as an intermediate medium between $\mathcal{S}$ and $\mathcal{R}$ with a view to ensuring a line-of-sight path between the two nodes. The unauthorized users known as eavesdroppers ($\mathcal{E}$) attempt to intercept the confidential information that is being transmitted from the $\mathcal{S}$ to $\mathcal{D}$. Based on the position of eavesdroppers, three different scenarios are considered where the eavesdropper attempts to overhear the communication.

\begin{itemize}
\item In \textit{Scenario-I}, the eavesdropper, $\mathcal{E}_{P}$, utilizes the RF link for wiretapping and both $\mathcal{R}$ and $\mathcal{E}_{P}$ attain analogous signal propagated from $\mathcal{I}_{P}$.

\item The \textit{Scenario-II} infers the eavesdropper, $\mathcal{E}_{Q}$, at the FSO link and both $\mathcal{D}$ and $\mathcal{E}_{Q}$ obtain the resembling propagated signal from $\mathcal{I}_{Q}$.

\item In \textit{Scenario-III}, the eavesdroppers  $\mathcal{E}_{P}$ and $\mathcal{E}_{Q}$ both attempt concurrently to overhear the confidential information from both the RF and FSO links.
\end{itemize}

\noindent
This model describes a passive eavesdropping scenario assuming that the RIS is oblivious to the CSI of eavesdroppers. Herein, $\mathcal{S}$  and $\mathcal{E}$ are equipped with a singular antenna, while $\mathcal{R}$ acts as a transceiver. $\mathcal{D}$ comprises a single photodetector for optical wave reception, while $\mathcal{I}_{P}$ and $\mathcal{I}_{Q}$ have $\mathcal{N}_{1}$ and $\mathcal{N}_{2}$ reflecting elements, respectively. The surface RF networks using $\mathcal{S}-\mathcal{I}_{P}-\mathcal{R}$ and $\mathcal{S}-\mathcal{I}_{P}-\mathcal{E}_{P}$ links pursue the Rician fading distribution. $\mathcal{R}$ serves to convert the obtained RF signal and redirect it as optical signal to $\mathcal{D}$ in the presence of $\mathcal{E}_{Q}$. Both the FSO links, $\mathcal{R}-\mathcal{I}_{Q}-\mathcal{D}$ and $\mathcal{R}-\mathcal{I}_{Q}-\mathcal{E}_{Q}$ experience Málaga turbulence with pointing error aided by a RIS, $\mathcal{I}_{Q}$.


\subsection{SNRs of Individual Links}

For \textit{Scenario-I}, $h_{{s_{p}}}$ ($s_{p}=1, 2, \ldots, {\mathcal{N}_1}$) indicates the first hop channel gain of both the $\mathcal{S}-\mathcal{I}_{P}-\mathcal{R}$ and $\mathcal{S}-\mathcal{I}_{P}-\mathcal{E}_{P}$ links. Similarly, $g_{{s_{p}}}$ and $n_{{s_{p}}}$ indicate the  channel gains of the later hop of those links, in a respective manner. Hence, the signals present at $\mathcal{R}$ and $\mathcal{E}_{P}$ are, correspondingly, represented as
\begin{align}
\label{r1}
y_{s,r}&=\left[\sum_{s_{p}=1}^{\mathcal{N}_1}
h_{{s_{p}}}e^{j\Phi_{s_{p}}}g_{{s_{p}}}\right]x+w_{1},
\\
\label{r2}
y_{s,e}&=\left[\sum_{s_{p}=1}^{\mathcal{N}_1}
h_{{s_{p}}}e^{j\Psi_{s_{p}}}n_{{s_{p}}}\right]x+w_{2}.
\end{align} 
With regards to these particular channels, we have $h_{{s_{p}}}=\alpha_{{s_{p}}}e^{j\varrho_{s_{p}}}$,  $g_{s_{p}}=\beta_{s_{p}}e^{j\vartheta_{s_{p}}}$, and {$n_{{s_{p}}}=\eta_{s_{p}}e^{j\delta_{s_{p}}}$}, where $\alpha_{s_{p}}$, $\beta_{s_{p}}$, and $\eta_{s_{p}}$ are the Rician distributed random variables (RVs), $\varrho_{{s_{p}}}$, $\vartheta_{s_{p}}$, and $\delta_{s_{p}}$ are the resembling phases of received signal gains, and $\Phi_{s_{p}}$ and $\Psi_{s_{p}}$ identifies the phase emanated by the $s_{p}$-th reflecting element of the RIS. In this work, we consider $\Phi_{s_{p}}\in[0,2\pi)$, $\Psi_{s_{p}}\in[0,2\pi)$, and the range of reflection on the assembled fortuitous signal present at the $s_{p}$-th element is deliberated as $1$. The conveyed data from $\mathcal{S}$ is represented in this scenario by $x$ with the power $S_s$ and $w_{1}\sim\mathcal{\widetilde{M}}(0,M_{r})$, $w_{2}\sim\mathcal{\widetilde{M}}(0,M_{e})$ are the additive white Gaussian noise (AWGN) samples with $M_r$, $M_e$ indicating the power of noise for the relevant networks. Mathematically, \eqref{r1} and \eqref{r2} are expressed as
\begin{align}
\label{r3}
y_{s,r}&=\textbf{g}^{T}\Phi\,\textbf{h}\,x+w_{1},
\\
\label{r4}
y_{s,e}&=\textbf{n}^{T}\Psi\,\textbf{h}\,x+w_{2},
\end{align} 
where the channel coefficient vectors are denoted by $\textbf{h}=[h_{1}\,h_{2}\,\ldots\,h_{\mathcal{N}_{1}}]^{T}$, $\textbf{g}=[g_{1}\,g_{2}\,\ldots\,g_{\mathcal{N}_{1}}]^{T}$, $\textbf{n}=[n_{1}\,n_{2}\,\ldots\,n_{\mathcal{N}_{1}}]^{T}$, and $\Phi=\text{diag}([e^{j\Phi_{1}}\,e^{j\Phi_{2}}\,\ldots\, e^{j\Phi_{\mathcal{N}_{1}}}])$ and $\Psi=\text{diag}([e^{j\Psi_{1}}\,e^{j\Psi_{2}}\,\ldots\, e^{j\Psi_{\mathcal{N}_{1}}}])$ are the diagonal matrices containing the transitions of phase employed by RIS components. The SNRs at $\mathcal{R}$ and $\mathcal{E}_{P}$ are expressed as
\begin{align}
\gamma_{\mathcal{R}}&=\frac{\left[\sum_{s_{p}=0}^{\mathcal{N}_{1}}\alpha_{{s_{p}}}
\beta_{s_{p}}e^{j\left(\Phi_{s_{p}}-\varrho_{s_{p}}-\vartheta_{s_{p}}\right)}\right]^{2} S_{s}}{M_{r}},
\\
\gamma_{\mathcal{E}_{P}}&=\frac{\left[\sum_{s_{p}=0}^{\mathcal{N}_{1}}\alpha_{s_{p}}\eta_{s_{p}}e^{j\left(\Psi_{s_{p}}-\varrho_{s_{p}}-{\delta_{s_{p}}}\right)}\right]^{2} S_{s}}{M_{e}}.
\end{align} 
It is worth noting that the ideal selection of $\Phi_{s_{p}}$ and $\Psi_{s_{p}}$ are $\Phi_{s_{p}}=\varrho_{s_{p}}+\vartheta_{s_{p}}$ and $\Psi_{s_{p}}=\varrho_{s_{p}}+\delta_{s_{p}}$ for obtaining maximized instantaneous SNR. Hence, the maximum possible SNRs at $\mathcal{R}$ and $\mathcal{E}_{P}$ are given, correspondingly, as
\begin{align}
\gamma_{\mathcal{R}}&=\left(\sum_{s_{p}=0}^{\mathcal{N}_{1}}\alpha_{{s_{p}}}
\beta_{s_{p}}\right)^{2} \gamma_{1},
\\
\gamma_{\mathcal{E}_{P}}&=\left(\sum_{s_{p}=0}^{\mathcal{N}_{1}}\alpha_{s_{p}}\eta_{s_{p}}\right)^{2} \gamma_{e_{1}},
\end{align} 
where the average SNR of the $\mathcal{S}-\mathcal{I}_{P}-\mathcal{R}$ link is denoted by $\gamma_{1}=\frac{S_{s}}{M_{r}}$ and the average SNR of $\mathcal{S}-\mathcal{I}_{P}-\mathcal{E}_{P}$ link is represented by $\gamma_{{e}_1}=\frac{S_{s}}{M_{e}}$ .

For \textit{Scenario-II}, the received signals at $\mathcal{D}$ and $\mathcal{E}_{Q}$ are represented in a form similar to the expressions in \eqref{r1}-\eqref{r4} and utilizing the same procedures for optimization, the received SNRs are given as
\begin{align}
\gamma_{\mathcal{D}}&=\left(\sum_{r_{q}=0}^{\mathcal{N}_{2}}\xi_{{r_{q}}}\beta_{r_{q}}\right)^{2}\gamma_{2},
\\
\gamma_{\mathcal{E}_{Q}}&=\left(\sum_{r_{q}=0}^{\mathcal{N}_{2}}\xi_{{r_{q}}}\eta_{r_{q}}\right)^{2}\gamma_{e_{2}},
\end{align}
where $\xi_{{r_{q}}}$, $\beta_{r_{q}}$, and $\eta_{r_{q}}$ are M{\'a}laga distributed RVs, $\gamma_{2}$ and $\gamma_{e_{2}}$ are the average SNRs of the $\mathcal{R}-\mathcal{I}_{Q}-\mathcal{D}$ and $\mathcal{R}-\mathcal{I}_{Q}-\mathcal{E}_{Q}$ links, respectively. 

The received SNR of the proposed literature that utilizes a variable gain AF relay is provided as 
\begin{align}
\gamma_{eq}&=\frac{\gamma_{\mathcal{R}}\,\gamma_{\mathcal{D}}}{\gamma_{\mathcal{R}}+\gamma_{\mathcal{D}}+1} \cong \min\left\{\gamma_{\mathcal{R}}, \gamma_{\mathcal{D}}\right\}.
\end{align} 
\subsection{PDF and CDF of $\gamma_{\mathcal{R}}$}

The probability density function (PDF) and CDF of $\gamma_{\mathcal{R}}$ are respectively expressed as \cite{salhab2021accurate}
\begin{align}
    \label{eq1}
    f_{\gamma_{\mathcal{R}}}(\gamma) &\simeq \frac{\gamma^{\frac{a-1}{2}} \exp \left(-\frac{\sqrt{\gamma}}{b \sqrt{{\gamma_1}}}\right)}{2 b^{a+1} \Gamma(a+1) {\gamma_1}^{\frac{a+1}{2}}}, 
\\
    \label{eq2}
F_{\gamma_{\mathcal{R}}}(\gamma)& \simeq \frac{\gamma\left(a+1, \frac{\sqrt{\gamma}}{b \sqrt{\gamma_1}}\right)}{\Gamma(a+1)},
\end{align} 
where $\gamma(\cdot, \cdot)$ is the lower incomplete Gamma function \cite[Eq.~(8.350.1)]{gradshteyn1988tables}, $\Gamma(\cdot)$ is the Gamma operator, $a$ and $b$ are constants related to the mean and variance of the cascaded Rician random variable ${\xi_s}_p$ computed as
\begin{align}
\label{eqa}
a=\frac{\mathbb{E}^2\left[{\xi_s}_p\right]}{\operatorname{Var}(\xi_m)}-1
\text{ and }
b=\frac{\operatorname{Var}(\xi_m)}{\mathbb{E}\left[{\xi_s}_p\right]},
\end{align} 
$\xi_m$ is the sum of i.i.d non-negative random variables,
\begin{align}
\label{eqa1}
\mathbb{E}\left[{\xi_s}_p\right]=\mathbb{E}\left[{\alpha_s}_p\right] \mathbb{E}\left[{\beta_s}_p\right],
\end{align} 
and $\mathbb{E}\left[{\alpha_s}_p\right]=\frac{1}{2} \sqrt{\frac{\pi \Omega_1}{K_1+1}} L_{1 / 2}\left(- K_1\right)$, $\mathbb{E}\left[{\beta_s}_p\right]=\frac{1}{2} \sqrt{\frac{\pi \Omega_2}{K_2+1}} L_{1 / 2}\left(- K_2\right)$, $\text{Var}(\xi_m)=\mathcal{N}_1 \text{Var}\left({\xi_s}_p\right)$, $K_{1}$ and $\Omega_{1}$ denote the shape parameter and scale parameter, respectively, for the first hop of $\mathcal{S}-\mathcal{I}_{P}-\mathcal{R}$ link, for the other hop those are expressed by $K_{2}$ and $\Omega_{2}$, correspondingly, $L_{1 / 2}(\cdot)$ denotes the Laguerre polynomial, i.e., $L_{1 / 2}(x)=$ $e^{x / 2}\left[(1-x) I_0\left(\frac{-x}{2}\right)-x I_1\left(\frac{-x}{2}\right)\right]$, and $I_v(\cdot)$ is the modified Bessel function of the first kind and order $v$ \cite[Eq.~(8.431)]{gradshteyn1988tables}. Further simplification of \eqref{eqa1} gives
\begin{align}
\mathbb{E}\left[{\xi_s}_p\right]&=\frac{\pi e^{-\frac{\left(K_1+K_2\right)}{2}}}{4} \sqrt{\frac{\Omega_1 \Omega_2}{\left(K_1+1\right)\left(K_2+1\right)}}\nonumber\\
 &\times\left[\left(K_1+1\right) I_0\left(\frac{K_1}{2}\right)+K_1 I_1\left(\frac{K_1}{2}\right)\right] \nonumber\\
 &\times\left[\left(K_2+1\right) I_0\left(\frac{K_2}{2}\right)+K_2 I_1\left(\frac{K_2}{2}\right)\right].
\end{align}
Notice that $\mathbb{E}\left[{\xi_s}_p^2\right]=\mathbb{E}\left[{\alpha_s}_p^2\right] \mathbb{E}\left[{\beta_s}_p^2\right]=\Omega_1 \Omega_2$.
The variance is computed as
\begin{align}
\operatorname{Var}\left({\xi_s}_p\right)=\mathbb{E}\left[{\xi_s}_p^2\right]-\mathbb{E}^2\left[{\xi_s}_p\right]=\Omega_1 \Omega_2-\mathbb{E}^2\left[{\xi_s}_p\right]. 
\end{align}

\subsection{PDF and CDF of $\gamma_{\mathcal{D}}$}
The PDF of $\gamma_{\mathcal{D}}$ can be calculated as \cite[Eq. (17)]{ndjiongue2021analysis}
\begin{align}
\label{eqfd}
f_{{\gamma}_{\mathcal{D}}}(\gamma)=\int_0^{\infty} f_{\gamma_h}(t) f_{\gamma_g}\left(\frac{\gamma}{t}\right) \frac{1}{t} dt.
\end{align}
The two sub-channels can be simulated by a unified distribution that takes into account pointing errors and turbulence levels assuming that the weather condition will remain the same throughout the environment \cite[Eq. (10)]{ansari2015performance}. The characterization of both sub-channels is done by $\alpha_h$, $m_h$, and $\xi_h$ for the $\mathcal{R}-\mathcal{I}_{Q}$ link and $\alpha_g$, $m_g$, and $\xi_g$ for the $\mathcal{I}_{Q}-\mathcal{D}$ link. The PDF of $\mathcal{R}-\mathcal{I}_{Q}$ and $\mathcal{I}_{Q}-\mathcal{D}$ links can be written as \cite[Eq. (10)]{ansari2015performance} 
\begin{align}
\label{eqpdf}
f_{\gamma_i}\left(\gamma_i\right)=\frac{\xi_i^2 {A_i}}{2^r \gamma_i} \sum_{m_i=1}^{\beta_i} b_m \mathrm{G}_{1,3}^{3,0}\left[{B_i}\left(\frac{\gamma_i}{{\bar {{\gamma}_{i}}}}\right)^{\frac{1}{r}} \biggl| \begin{array}{c}
{\xi_i}^2+1 \\
{\xi_i}^2, {\alpha_i}, m_i
\end{array}\right],
\end{align}
where $i \in\{h, g\}$, $ A_i \triangleq \frac{2 {\alpha_i}^{\alpha_i / 2}}{c^{1+\alpha_i / 2} \Gamma(\alpha_i)}\left(\frac{c\beta_i}{c\beta_i+\Omega^{\prime}}\right)^{\beta_i+\alpha_i / 2}$, 
$ a_{m_{i}} \triangleq\left(\begin{array}{c}
\beta_i-1 \\
{m_{i}}-1
\end{array}\right) \frac{\left(c \beta_i+\Omega^{\prime}\right)^{1-{m_{i}} / 2}}{({m_{i}}-1) !}\left(\frac{\Omega^{\prime}}{c}\right)^{{m_{i}}-1}\left(\frac{\alpha_i}{\beta_i}\right)^{{m_{i}} / 2}$, ${b_m}_i={a_m}_i[\alpha_i \beta_i /(c \beta_i+\Omega^{\prime})]^{-(\alpha_i+m_i) / 2}$,  $B_i=\xi_i^2 \alpha_i \beta_i(c+\Omega^{\prime}t) /[(\xi_i^2+1)(c \beta_i+\Omega^{\prime})]$, $\Omega^{\prime}=\Omega+2 b_0 \rho+$ $2 \sqrt{2 b_0 \rho \Omega} \cos(\phi_A-\phi_B)$, $c=2b_0\left(1-\rho\right)$ indicates the average amount of power received by off-axis eddies from the dispersive element, {$\alpha_i$ and $\beta_i$ are the turbulence parameters, $\bar{\gamma}_i$ is the average SNR,  $\xi_i$ represents pointing error, $\Omega$ is the average power of LOS component, $b_0$ is the average power of the total scatter components, $\rho$ represents the quantity of scattering power coupled to the LOS component
, $\phi_{A}$ and $\phi_{B}$ are the deterministic phases
of the LOS and the coupled-to-LOS scatter terms, respectively, $ r \in\{1,2\}$ determines if the transmission makes use of the heterodyne detection $(\mathrm{HD})$ approach $(r=1)$ or the intensity modulation/direct detection $(\mathrm{IM/DD})$ techniques ($r=2$) \cite{ansari2015performance}, and $\mathbf{G}_{p, q}^{m, n}\left[z \mid \begin{array}{l}a_p \\ b_q\end{array}\right]$ is the Meijer's G function \cite[Eq.~(9.301)]{gradshteyn1988tables}. We sequentially substitute $\gamma_i$ by t and ${\frac{\gamma}{t}}$ in (\ref{eqpdf}), and obtain $f_{\gamma_h}\left(t\right)$ and $f_{\gamma_g}\left({\frac{\gamma}{t}}\right)$, respectively, as
\begin{align}
\label{eqfh}
&f_{\gamma_h}\left(t\right)=\frac{\xi_h^2 {A_h}}{2^r t} \sum_{m_h=1}^{\beta_h} b_{{m}_h}\,G_{1,3}^{3,0}\left[{B_h}\left(\frac{t}{\bar {\gamma}_{h}}\right) ^{\frac{1}{r}} \biggl|
\begin{array}{c}
{\xi_h}^2+1 \\
{\xi_h}^2, {\alpha_h}, m_h
\end{array}\right],
\\
\label{eq34}
&f_{\gamma_g}\left({\frac{\gamma}{t}}\right)=\frac{\xi_g^2 {A_g}t}{2^r \gamma}\, \sum_{m_g=1}^{\beta_g} b_{{m}_g}
G_{1,3}^{3,0}\left[{B_g}\left(\frac{\gamma}{t{\bar {\gamma}_{g}}}\right)^{\frac{1}{r}}\biggl| \begin{array}{c}
{\xi_g}^2+1 \\
{\xi_g}^2, {\alpha_g}, m_g
\end{array}\right],
\end{align}
where ${\bar {\gamma}_{h}}$ and ${\bar {\gamma}_{g}}$ are the average SNRs. In \eqref{eq34}, the variable $t$ appears in the denominator. Utilizing the Meijer's G function's reflection characteristic \cite{karp2016hypergeometric} in \eqref{eq34}, we are able to obtain a Meijer's G function that has a numerator-based variable named $t$ as
\begin{align}
\label{eqfg}
f_{\gamma_g}\left({\frac{\gamma}{t}}\right)&=\frac{\xi_g^2 {A_g}t}{2^r \gamma} \sum_{m_g=1}^{\beta_g} b_{{m}_g} \nonumber
\\ 
& \times G_{3,1}^{0,3}\left[\frac{1}{B_g}\left(\frac{{t\bar{\gamma_g}}}{\gamma}\right)^{\frac{1}{r}} \biggl| \begin{array}{c}
1-{\xi_g}^2,1- {\alpha_g},1-m_g\\
{-\xi_g}^2
\end{array}\right].
\end{align}
Substituting \eqref{eqfh} and \eqref{eqfg} into \eqref{eqfd} then applying the change of variable $X=t^{\frac{1}{a}} \Rightarrow t=X^a$ and $d t=$ $a X^{a-1} d X$  via utilizing \cite[Eq. (2.24.1.1)]{brychkov1986integrals}, we obtain the exact unified PDF of end-to-end SNR as
\begin{align}
\label{eqfdf}
f_{{\gamma}_{\mathcal{D}}}(\gamma)&=\frac{\xi_h^2 {A_h}\xi_g^2 {A_g}r}{2^{2r} \gamma} \sum_{m_h=1}^{\beta_h}\sum_{m_g=1}^{\beta_g} b_{{m}_h} b_{{m}_g} \nonumber \\ &\times G_{2,6}^{6,0}\left[{B_h}{B_g}\left(\frac{\gamma}{{\gamma_2}}\right)^{\frac{1}{r}} \biggl| \begin{array}{c}
{1+\xi_g}^2, 1+\xi_h^2 \\
{\xi_h}^2, {\alpha_h}, m_h,\xi_g^2, \alpha_g, m_g
\end{array}\right],
\end{align}
where ${\gamma_2}$= $\bar{\gamma}_h\bar{\gamma}_g$.
The CDF of the end-to-end SNR can be written as
\begin{align}
F_{{\gamma}_{\mathcal{D}}}(\gamma)=\int_0^{\gamma} f_{{\gamma}_{\mathcal{D}}}(\gamma) d\gamma,
\label{eqn:cdffso}
\end{align}
By substituting \eqref{eqfdf} into \eqref{eqn:cdffso}, we obtain the CDF of $\gamma_{\mathcal{D}}$  via utilizing \cite[Eq. (07.34.21.0084.01)]{research2010mathematica} as
\begin{align}
\label{cdffsov}
F_{\gamma_{\mathcal{D}}}(\gamma)&= \frac{\xi_h^2 {A_h}\xi_g^2 {A_g}}{2^{2r} }\sum_{m_h=1}^{\beta_h}\sum_{m_g=1}^{\beta_g}b_{{m}_h} b_{{m}_g} \frac{{r}^{\alpha_h+\alpha_g+m_h+m_g-2}}{{2\pi}^{2(r-1)}} \nonumber \\ &\times G_{2r+1,6r+1}^{6r,1}\left[\left(\frac{{B}_{e q}}{\gamma_2}\right)\gamma \biggl| \begin{array}{c}
{1,l_{{g}_1}, l_{{h}_1}} \\
l_{{h}_2}, l_{{g}_2}, 0
\end{array}\right],
\end{align}
where ${B}_{eq}=\frac{\left({B_h}{B_g}\right)^r}{r^{4r}},  
l_{g_1}=\left(\frac{1+{\xi_g}^2}{r},\ldots, \frac{1+{\xi_g}^2+r-1}{r}\right), 
l_{h_1}=\left(\frac{1+{\xi_h}^2}{r}, \ldots, \frac{1+{\xi_h}^2+r-1}{r}\right)$,
$ l_{g_2}=\biggl(\frac{{\xi_g}^2}{r},  \ldots, \frac{{\xi_g}^2+r-1}{r}$, $\frac{\alpha_g}{r},\ldots, \frac{\alpha_g+r-1}{r},\frac{m_g}{r}, \ldots, \frac{m_g+r-1}{r}\biggl)$, and
$ l_{h_2}=\left(\frac{{\xi_h}^2}{r},  \ldots, \frac{{\xi_h}^2+r-1}{r}, \frac{\alpha_h}{r}, \ldots, \frac{\alpha_h+r-1}{r},\frac{m_h}{r}, \ldots, \frac{m_h+r-1}{r}\right)$.

\subsection{PDF of $\gamma_{\mathcal{E}}$}

Presuming $\mathcal{S}-\mathcal{I}_{\mathcal{P}}-\mathcal{E}_{\mathcal{P}}$ link (\textit{Scenario-I}) also considers Rician distribution, the PDF and CDF of $\gamma_{\mathcal{E}_P}$ are expressed as 
\begin{align}
\label{pdfev1}
f_{{\gamma}_{\mathcal{E}_{\mathcal{P}}}}(\gamma) &\simeq \frac{\gamma^{\frac{a_e-1}{2}} \exp \left(-\frac{\sqrt{\gamma}}{b_e \sqrt{\gamma_{{e}_1}}}\right)}{2 b_e^{a_e+1} \Gamma(a_e+1) {\gamma_{{e}_1}}^{\frac{a_e+1}{2}}},
\\
\label{cdfev1}
F_{{\gamma}_{\mathcal{E}_{\mathcal{P}}}}(\gamma) &\simeq \frac{\gamma\left(a_e+1, \frac{\sqrt{\gamma}}{b_e \sqrt{\gamma_{e_{1}}}}\right)}{\Gamma(a_e+1)},
\end{align} 
where $a_e$ and $b_e$ are constants that can be expressed in the respective manner as \eqref{eqa}
\begin{align}
a_e=\frac{\mathbb{E}^2\left[{\xi_p}_e\right]}{\operatorname{Var}(\xi_e)}-1
\text{ and }
\label{eqbe}
b_e=\frac{\operatorname{Var}(\xi_e)}{\mathbb{E}\left[{\xi_p}_e\right]}.
\end{align}
Similar to \eqref{eqa1}
\begin{align}
\label{eq33}
\mathbb{E}\left[{\xi_p}_e\right]=\mathbb{E}\left[\alpha_{s_{p}}\right] \mathbb{E}\left[\eta_{s_{p}}\right],
\end{align}
where $\mathbb{E}\left[\eta_{s_{p}}\right]=\frac{1}{2} \sqrt{\frac{\pi \Omega_3}{K_3+1}} L_{1 / 2}\left(-K_3\right)$, ${\operatorname{Var}(\xi_e)}=\mathcal{N}_1 \operatorname{Var}\left({\xi_p}_e\right)$, and $K_{3}$ and $\Omega_{3}$ are the shape and scale parameters for the second hop of $\mathcal{S}-\mathcal{I}_{\mathcal{P}}-\mathcal{E}_{\mathcal{P}}$ link. Further simplification of \eqref{eq33} gives
\begin{align}
\mathbb{E}\left[{\xi_p}_e\right]&= \frac{\pi e^{-\frac{\left(K_1+K_3\right)}{2}}}{4} \sqrt{\frac{\Omega_1 \Omega_3}{\left(K_1+1\right)\left(K_3+1\right)}}\nonumber\\
 &\times\left[\left(K_1+1\right) I_0\left(\frac{K_1}{2}\right)+K_1 I_1\left(\frac{K_1}{2}\right)\right] \nonumber\\
 &\times\left[\left(K_3+1\right) I_0\left(\frac{K_3}{2}\right)+K_3 I_1\left(\frac{K_3}{2}\right)\right].
\end{align}
Notice that $\mathbb{E}\left[{\xi_p}_e^2\right]=\mathbb{E}\left[{\alpha_s}_p^2\right] \mathbb{E}\left[{\eta_s}_p^2\right]=\Omega_1 \Omega_3$. The variance is computed as
\begin{align}
\operatorname{Var}\left({\xi_p}_e\right)=\mathbb{E}\left[{\xi_p}_e^2\right]-\mathbb{E}^2\left[{\xi_p}_e\right]=\Omega_1 \Omega_3-\mathbb{E}^2\left[{\xi_p}_e\right].
\end{align}

For \textit{Scenario-II}, considering $\mathcal{R}-\mathcal{I}_{\mathcal{Q}}-\mathcal{E}_{\mathcal{Q}}$ link experiences M{\'a}laga distribution, the PDF of $\gamma_{\mathcal{E}_{\mathcal{Q}}}$ is expressed as
\begin{align}
\label{pdfev2}
f_{{\gamma}_{\mathcal{E}_{\mathcal{Q}}}}(\gamma)&=\frac{{\xi_h}^2 {{A_h}}{\xi_g}_e^2 {{A_g}_e}}{2^{2r} \gamma} \sum_{{m_h}=1}^{{\beta_h}}\sum_{{m_g}_e=1}^{{\beta_g}_e} b_{{m}_{h}} b_{{m}_{{g}_e}}\nonumber \\ &\times G_{2,6}^{6,0}\left[V_1\gamma^{\frac{1}{r}} \biggl| \begin{array}{c}
{1+{\xi_g}_e}^2, 1+{\xi_h}^2 \\
{\xi_h}^2, {\alpha_h}, {m_h},{\xi_g}_e^2, {\alpha_g}_e, {m_g}_e
\end{array}\right],
\end{align}
where the sub-channel link $\mathcal{I}_{Q}-\mathcal{{E}_Q}$ is characterized by $\alpha_{{g}_e}$, $m_{g_{e}}$ and $\xi_{{g}_e}$, $ A_{g_{e}} \triangleq \frac{2 {\alpha_{g_{e}}}^{\alpha_{g_{e}} / 2}}{c^{1+\alpha_{g_{e}} / 2} \Gamma(\alpha_{g_{e}})}(\frac{c \beta_{g_{e}}}{c \beta_{g_{e}}+\Omega^{\prime}})^{\beta_{g_{e}}+\alpha_{g_{e}} / 2}$, ${b_m}_{g_{e}}={a_m}_{g_{e}}[\alpha_{g_{e}} \beta_{g_{e}} /(c\beta_{g_{e}}+\Omega^{\prime})]^{-(\alpha_{g_{e}}+m_{g_{e}}) / 2}$, ${a_m}_{g_{e}} \triangleq\left(\begin{array}{c}
\beta_i-1 \\
m_{g_{e}}-1
\end{array}\right) \frac{\left(c \beta_i+\Omega^{\prime}\right)^{1-{m_{g_{e}}} / 2}}{({m_{g_{e}}}-1) !}\left(\frac{\Omega^{\prime}}{c}\right)^{{m_{g_{e}}}-1}\left(\frac{\alpha_i}{\beta_i}\right)^{{m_{g_{e}}} / 2}$, $B_{g_{e}}=\xi_{g_{e}}^2 \alpha_{g_{e}} \beta_{g_{e}}(c+\Omega^{\prime}t) /[(\xi_{g_{e}}^2+1)(c \beta_{g_{e}}+\Omega^{\prime})]$, and 
$V_1=B_{{h}}B_{{g}_e}\left(\frac{1}{\gamma_{{e}_2}}\right)^{\frac{1}{r}}$.
\subsection{CDF of End-to-End SNR for RIS-aided Dual-hop RF-FSO Link}

The CDF of $\gamma_{eq}$ is expressed as 
\begin{align}
\label{eq9}
F_{\gamma_{eq}}(\gamma)=F_{\gamma_{\mathcal{R}}}(\gamma)+F_{\gamma_{\mathcal{D}}}(\gamma)-F_{\gamma_{\mathcal{R}}}(\gamma)F_{\gamma_{\mathcal{D}}}(\gamma).
\end{align} 
Substituting \eqref{eq2} and \eqref{cdffsov} in \eqref{eq9} and carrying out algebraic calculations
, the simplification of CDF of $\gamma_{eq}$ can be attained as
\begin{align}
\label{cdfeveq}
\nonumber
F_{\gamma_{eq}}(\gamma)&=\sum_{n=0}^{\infty} \frac{(-1)^n\left(\frac{1}{b \sqrt{{\gamma}_1}}\right)^{a+1+n}}{n !\left(a+1+n\right)\Gamma \left(a+1\right)} + \frac{\xi_h^2 {A_h}\xi_g^2 {A_g}}{2^{2r} }\\ & \times \sum_{m_h=1}^{\beta_h} 
\sum_{m_g=1}^{\beta_g}b_{{m}_h} b_{{m}_g} \frac{{r}^{\alpha_h+\alpha_g+m_h+m_g-2}}{{2\pi}^{2(r-1)}} \nonumber \\ & \times G_{2r+1,6r+1}^{6r,1}\left[\left(\frac{{B}_{e q}}{{\gamma_2}}\right)\gamma \biggl| \begin{array}{c}
{1,l_{{g}_1}, l_{{h}_1}} \\
l_{{h}_2}, l_{{g}_2}, 0
\end{array}\right] \nonumber\\ & \times \left(1-\sum_{n=0}^{\infty} \frac{(-1)^n\left(\frac{1}{b \sqrt{{\gamma}_1}}\right)^{a+1+n}}{n !\left(a+1+n\right)\Gamma \left(a+1\right)} \right).
\end{align}
As per the comprehension discussed in the literature review section, the combination of RIS-assisted RF-FSO framework taking Rician and M{\'a}laga distributions into account has not yet been described in any current study within literature. As a result, the expression found in \eqref{cdfeveq} can be demonstrated to be unique.
 Also, the generalized depiction of both Rician and M{\'a}laga distribution drives this endeavor towards the goal of unifying the many existing models by treating them as prominent occurrences.

\section{{\color{black}Performance Analysis}}
{\color{black}In this portion of the work, we attain the expressions for the suggested RIS-assisted RF-FSO network's metrics of performance, namely ASC, the lower bound of SOP, SPSC, IP, and EST.

\subsection{Average Secrecy Capacity Analysis}

ASC is the mean value of the instantaneous secrecy capacity, which can be stated analytically as \cite[Eq.~(15)]{islam2020secrecy}
\begin{align}
\label{eq11}
ASC^{I}=\int_{0}^{\infty}\frac{1}{1+\gamma}\,F_{{\gamma}_{\mathcal{E}_{\mathcal{P}}}}(\gamma)\,\left[1-F_{\gamma_{eq}}(\gamma)\right]\,d\gamma.
\end{align} 
On substituting \eqref{cdfev1} and \eqref{cdfeveq} into \eqref{eq11}, ASC is derived as
\begin{align}
\nonumber
ASC^{I}&= {\mathcal{X}_1}\biggl(\sum_{n=0}^{\infty}{\mathcal{U}_1} -\sum_{n=0}^{\infty}{\mathcal{X}_2}{\mathcal{U}_2}-\sum_{m_h=1}^{\beta_h} 
\sum_{m_g=1}^{\beta_g} {\mathcal{X}_3}{\mathcal{U}_3}
\\
& +\sum_{n=0}^{\infty} \sum_{m_h=1}^{\beta_h} 
\sum_{m_g=1}^{\beta_g} {\mathcal{X}_2}{\mathcal{X}_3} {\mathcal{U}_4}\biggl),
\end{align}
where ${\mathcal{X}_1}= \frac{(-1)^n\left(\frac{1}{b_e \sqrt{{\gamma}_1}}\right)^{a_e+n+1}}{n !\left(a_e+n+1\right)\Gamma \left(a_e+1\right)},{\mathcal{X}_2}= \frac{(-1)^n\left(\frac{1}{b \sqrt{{\gamma}_1}}\right)^{a+n+1}}{n !\left(a+n+1\right)\Gamma \left(a+1\right)},{\mathcal{X}_3}=b_{{m}_h} b_{{m}_g} \frac{{r}^{\alpha_h+\alpha_g+m_h+m_g-2}}{{2\pi}^{2(r-1)}}$ and four integral expressions ${\mathcal{U}_1}$, ${\mathcal{U}_2}$, ${\mathcal{U}_3}$ and ${\mathcal{U}_4}$ are expressed as follows.
\subsubsection{Derivation of ${\mathcal{U}_1}$}

${\mathcal{U}_1}$ is expressed as
\begin{align}
{\mathcal{U}_1}=\int_0^{\infty}\frac{{\gamma}^{\frac{a_e+n+1}{2}}}{1+\gamma} d\gamma.
\end{align}
With the fulfillment of identity \cite[Eq.~(3.194.3)]{gradshteyn1988tables}, ${\mathcal{U}_1}$ is attained as
\begin{align}
{\mathcal{U}_1}=\mathcal{B}\left(\frac{a_e+n+3}{2},1-\frac{a_e+n+3}{2}\right),
\end{align} 
where $\mathcal{B}\left(.,.\right)$ is the Beta function \cite[Eq.~(8.39)]{gradshteyn1988tables}.

\subsubsection{Derivation of ${\mathcal{U}_2}$}

${\mathcal{U}_2}$ is expressed as
\begin{align}
{\mathcal{U}_2}=\int_0^{\infty}\frac{{\gamma}^{\frac{a+a_e+2n+2}{2}}}{1+\gamma} d\gamma.
\end{align} 
Using an analogous method to the one used to derive ${\mathcal{U}_1}$, ${\mathcal{U}_2}$ is closed in as
\begin{align}
{\mathcal{U}_2}=\mathcal{B}\left(\frac{a+a_e+2n+4}{2},1-\frac{a+a_e+2n+4}{2}\right).
\end{align}

\subsubsection{Derivation of ${\mathcal{U}_3}$}

${\mathcal{U}_3}$ is expressed as
\begin{align}
{\mathcal{U}_3}=\int_0^{\infty}\frac{{\gamma}^{\frac{a_e+n+1}{2}}}{1+\gamma} G_{2r+1,6r+1}^{6r,1}\left[\left(\frac{{B}_{e q}}{{\gamma_2}}\right)\gamma \biggl| \begin{array}{c}
{1,l_{{g}_1}, l_{{h}_1}} \\
l_{{h}_2}, l_{{g}_2}, 0
\end{array}\right] d\gamma.
\end{align} 
The transformation of $\frac{1}{1+\gamma}$ into Meijer's G function is done by utilizing the identity \cite[Eq.~(8.4.2.5)]{brychkov1986integrals} and solving the integral upon utilization of \cite[Eq.~(2.24.1.1)]{brychkov1986integrals}, ${\mathcal{U}_3}$ is obtained as
\begin{align}
{\mathcal{U}_3}&=\int_0^{\infty}{\gamma}^{\frac{a_e+n+1}{2}} G_{2r+1,6r+1}^{6r,1}\left[\left(\frac{{B}_{e q}}{{\gamma_2}}\right)\gamma \biggl| \begin{array}{c}
{1,l_{{g}_1}, l_{{h}_1}} \\
l_{{h}_2}, l_{{g}_2}, 0
\end{array}\right] \nonumber \\ &\times G_{1,1}^{1,1}\left[\gamma \biggl| \begin{array}{c}
{0} \\
 0
\end{array}\right]d\gamma=\left(\frac{{B}_{e q}}{{\gamma_2}}\right)^{-\alpha_1}\nonumber \\
& \times G_{6r+2,2r+2}^{2,6r+1}\left[\frac{{\gamma_2}}{{B}_{e q}} \biggl| \begin{array}{c}
{0,-\alpha_1-l_{{h}_2},-\alpha_1-l_{{g}_2},-\alpha_1} \\
0, -\alpha_1-1,-\alpha_1-l_{{g}_1},-\alpha_1-l_{{h}_1}
\end{array}\right],
\end{align}
where $\alpha_1=\frac{a_e+n+3}{2}$.

\subsubsection{Derivation of ${\mathcal{U}_4}$}

${\mathcal{U}_4}$ is expressed as
\begin{align}
{\mathcal{U}_4}=\int_0^{\infty}\frac{{\gamma}^{\frac{a+a_e+2n+2}{2}}}{1+\gamma} G_{2r+1,6r+1}^{6r,1}\left[\left(\frac{{B}_{e q}}{{\gamma_{2}}}\right)\gamma \biggl| \begin{array}{c}
{1,l_{{g}_1}, l_{{h}_1}} \\
l_{{h}_2}, l_{{g}_2}, 0
\end{array}\right] d\gamma.
\end{align}
Using an analogous method to the one used to derive ${\mathcal{U}_3}$, ${\mathcal{U}_4}$ is derived as
\begin{align}
{\mathcal{U}_4}&=\int_0^{\infty}{\gamma}^{\frac{a_e+n+1}{2}} G_{2r+1,6r+1}^{6r,1}\left[\left(\frac{{B}_{e q}}{{\gamma_2}}\right)\gamma \biggl| \begin{array}{c}
{1,l_{{g}_1}, l_{{h}_1}} \\
l_{{h}_2}, l_{{g}_2}, 0
\end{array}\right] \nonumber \\ &\times G_{1,1}^{1,1}\left[\gamma \biggl| \begin{array}{c}
{0} \\
 0
\end{array}\right]d\gamma=\left(\frac{{B}_{e q}}{{\gamma_2}}\right)^{-\alpha_2-1}\nonumber \\
& \times G_{6r+2,2r+2}^{2,6r+1}\left[\frac{{\gamma_2}}{{B}_{e q}} \biggl| \begin{array}{c}
{0,-\alpha_2-l_{{h}_2},-\alpha_2-l_{{g}_2},-\alpha_2} \\
0, -\alpha_2-1,-\alpha_2-l_{{g}_1},-\alpha_2-l_{{h}_1}
\end{array}\right],
\end{align}
where $\alpha_2=\frac{a+a_e+2n+2}{2}$.


\subsection{Lower Bound of Secrecy Outage Probability Analysis}
\textbf{\textit{Scenario-I:}}
In accordance with \cite[Eq.~(21)]{sarker2020secrecy}, the lower bound of SOP can be stated as
\begin{align}
\label{sopev1}
SOP^{I}=\text{Pr}\left\{\gamma_{eq}\leq\phi\,\gamma_{\mathcal{E}}\right\}=\int_{0}^{\infty}F_{\gamma_{eq}}(\phi\,\gamma)\,f_{\gamma_{\mathcal{E}_P}}(\gamma)\,d\gamma,
\end{align}
where $\phi=2^{R_{s}}$. Now, substituting \eqref{pdfev1} and \eqref{cdfeveq} into \eqref{sopev1}, SOP is expressed finally as
\begin{align}
\label{eqn:sop3}
SOP^{I}&=\frac{{{\mathcal{M}_1} {\mathcal{R}_1}+{\mathcal{M}_2}{\mathcal{R}_2}-{\mathcal{M}_3}{\mathcal{R}_3}}}{2 b_e^{a_e+1} \Gamma(a_e+1) {\gamma_{{e}_1}}^{\frac{a_e+1}{2}}},
\end{align}
where ${\mathcal{M}_1}=\sum_{n=0}^{\infty} \frac{(-1)^n}{n !\left(a+1+n\right)\Gamma(a+1)}\left(\frac{\phi^\frac{1}{2}}{b \sqrt{{\gamma}_1}}\right)^{a+1+n}$, $ {\mathcal{M}_2}=\frac{\xi_h^2 {A_h}\xi_g^2 {A_g}}{2^{2r} }\sum_{m_h=1}^{\beta_h}\sum_{m_g=1}^{\beta_g}{b_{m_h}} {b_{m_g}} \frac{{r}^{\alpha_h+\alpha_g+m_h+m_g-2}}{{2\pi}^{2(r-1)}}$, ${\mathcal{M}_3}=\frac{{\mathcal{M}_1} \xi_h^2 \xi_g^2} {2^{2r} }{A_h}{A_g}\sum_{m_h=1}^{\beta_h}\sum_{m_g=1}^{\beta_g}\frac{{r}^{\alpha_h+\alpha_g+m_h+m_g-2}}{{2\pi}^{2(r-1)}}{b_{m_h}} {b_{m_g}} $ and derivations of the three integral terms ${\mathcal{R}_1}$, ${\mathcal{R}_2}$ and ${\mathcal{R}_3}$ are expressed as follows.

\subsubsection{Derivation of ${\mathcal{R}_1}$}

${\mathcal{R}_1}$ is expressed as
\begin{align}
{\mathcal{R}_1}=\int_0^{\infty}{\gamma}^{\frac{a+a_e+n}{2}}e^-{\frac{\sqrt{\gamma}}{b_e\sqrt{{\gamma_{{e}_1}}}} } d\gamma.\nonumber
\end{align}
${\mathcal{R}_1}$ is derived by utilizing \cite[Eq. (3.326.2)]{gradshteyn1988tables} as
\begin{align}
{\mathcal{R}_1}=2(a+a_e+n+1)!\left(\frac{1}{b_e\sqrt{{\gamma_{{e}_1}}}}\right)^{-a-a_e-n-2}.
\end{align}

\subsubsection{Derivation of ${\mathcal{R}_2}$}

${\mathcal{R}_2}$ is expressed as
\begin{align}
{\mathcal{R}_2}&=\int_0^{\infty}{\gamma}^{\frac{a_e+1}{2}-1} G_{2r+1,6r+1}^{6r,1}\left[\left(\frac{{B}_{e q}\phi}{{\gamma_2}}\right)\gamma \biggl| \begin{array}{c}
{1,l_{g_1}, l_{h_1}} \\
l_{h_2}, l_{g_2}, 0
\end{array}\right] \nonumber \\
& \times e^{\frac{-\sqrt{\gamma}}{b_e\sqrt{{\gamma_{{e}_1}}}} } d{\gamma}\nonumber
\end{align}
Now, through the use of a number of mathematical operations utilizing \cite[Eq.~(8.4.3.1) and (2.24.1.1)]{brychkov1986integrals}, ${\mathcal{R}_2}$ is derived as
\begin{align}
{\mathcal{R}_2}&=\int_0^{\infty}{\gamma}^{\frac{a_e+1}{2}-1} G_{2r+1,6r+1}^{6r,1}\left[\left(\frac{{B}_{e q}\phi}{{\gamma_2}}\right)\gamma \biggl| \begin{array}{c}
{1,l_{g_1}, l_{h_1}} \\
l_{h_2}, l_{g_2}, 0
\end{array}\right] \nonumber \\
& \times G_{0,1}^{1,0}\left[\left(\frac{\sqrt{\gamma}}{b_e\sqrt{{\gamma_{{e}_1}}}}\right) \biggl| \begin{array}{c}
{-} \\
 0
\end{array}\right]d{\gamma}={\mathcal{Z}_1}\nonumber\\
&  \times G_{6r+1,2r+3}^{3,6r}\left[\frac{\left(\frac{1}{b_e \sqrt{{\gamma_{{e}_1}}}}\right)^2}{4B_{e q}\phi}\gamma_2 \biggl| \begin{array}{c}
{l_{h_3},l_{g_3}, 1-\frac{a_e+1}{2}} \\
0,-\frac{a_e+1}{2},l_{g_4}, l_{h_4}, 0
\end{array}\right],
\end{align}
where ${\mathcal{Z}_1}={\pi}^{-\frac{1}{2}} \left(\frac{{\gamma_2}}{B_{e q}\phi}\right)^{\frac{a_e+1}{2}}, l_{h_3}=1-\frac{a_e+1}{2}-l_{h_2}, l_{g_3}=1-\frac{a_e+1}{2}-l_{g_2}$, $l_{h_4}=1-\frac{a_e+1}{2}-l_{h_1}$, and $l_{g_4}=1-\frac{a_e+1}{2}-l_{g_1}$.
\subsubsection{Derivation of ${\mathcal{R}_3}$}

${\mathcal{R}_3}$ is expressed as
\begin{align}
\nonumber
{\mathcal{R}_3}&=\int_0^{\infty}{\gamma}^{\frac{a+a_e+n}{2}}\mathrm{G}_{2r+1,6r+1}^{6r,1}\left[\left(\frac{{B}_{e q}\phi}{{\gamma_2}}\right)\gamma \biggl| \begin{array}{c}
{1,l_{g_1}, l_{h_1}} \\
l_{h_2}, l_{g_2}, 0
\end{array}\right] \\ &
\times e^{\frac{-\sqrt{\gamma}}{b_e\sqrt{{\gamma_{{e}_1}}}} }d{\gamma}.\nonumber
\end{align}
Using an analogous method to the one used to derive ${\mathcal{R}_2}$, ${\mathcal{R}_3}$ is derived as
\begin{align}
{\mathcal{R}_3}&=\int_0^{\infty}{\gamma}^{\frac{a+a_e+n}{2}}\mathrm{G}_{2r+1,6r+1}^{6r,1}\left[\left(\frac{{B}_{e q}\phi}{{\gamma_2}}\right)\gamma \biggl| \begin{array}{c}
{1,l_{g_1}, l_{h_1}} \\
l_{h_2}, l_{g_2}, 0
\end{array}\right] \nonumber \\&
\times \mathrm{G}_{0,1}^{1,0}\left[\left(\frac{\sqrt{\gamma}}{b_e \sqrt{{\gamma_{{e}_1}}}}\right) \biggl| \begin{array}{c}
{-} \\
 0
\end{array}\right]d{\gamma}\nonumber\\
&={\mathcal{Z}_2} \mathrm{G}_{6r+1,2r+3}^{3,6r}\left[\frac{\left(\frac{1}{b_e\sqrt{{\gamma_{{e}_1}}}}\right)^2}{4B_{e q}\phi}\gamma_2 \biggl| \begin{array}{c}
{l_{h_5},l_{g_5}, 1-{\mathcal{Z}_3}} \\
0,- {\mathcal{Z}_3},l_{g_6}, l_{h_6}
\end{array}\right],
\end{align}
where ${\mathcal{Z}_2}={\pi}^{-\frac{1}{2}} \left(\frac{{\gamma_2}}{B_{e q}\phi}\right)^{\mathcal{Z}_3}$, $ {\mathcal{Z}_3}=\frac{a+a_e+n+2}{2}$, $l_{h_5}=1-{\mathcal{Z}_3}-l_{h_2}, l_{g_5}=1-{\mathcal{Z}_3}-l_{g_2}, l_{h_6}=1-{\mathcal{Z}_3}-l_{h_1}$, and $ l_{g_6}=1-{\mathcal{Z}_3}-l_{g_1}$.

\textbf{\textit{Scenario-II:}}
The expression of SOP of RIS-aided combined RF-FSO channel when the eavesdropper is at the FSO link can be described as \cite[Eq.~(13)]{sarker2021intercept}
\begin{align}
{SOP}^{II} &=\operatorname{Pr}\left\{C_{s c} \leq R_{s }\right\} =\operatorname{Pr}\left\{\gamma_{e q} \leq \phi \gamma_{\mathcal{E}_Q}+\phi-1\right\} \nonumber\\
&=\int_0^{\infty} \int_{\phi \gamma+\phi-1}^{\infty} f_{\gamma_{e q}}\left(\gamma\right) f_{\gamma_{\mathcal{E}_{\mathcal{Q}}}}\left(\gamma\right) d \gamma_{e q} d \gamma\nonumber \\
&=\int_0^{\infty} F_{{\gamma}_{\mathcal{D}}}\left(\phi \gamma+\phi-1\right) f_{\gamma_{\mathcal{E}_{\mathcal{Q}}}}\left(\gamma\right) d \gamma \nonumber \\ & \times  (1-F_{{\gamma}_{\mathcal{R}}}(\phi-1))+F_{{\gamma}_{\mathcal{R}}}(\phi-1).
\end{align}
Due to mathematical process of obtaining the precise $\mathrm{SOP}$ closed-form formula being challenging, we deduce the mathematical expression of $\mathrm{SOP}$ at the lower-bound considering the variable gain relaying scheme as \cite[Eq.~(14)]{sarker2021intercept}
\begin{align}
\label{eqn:soplower2}
{SOP}^{II}\geq {SOP}^{L} &=\operatorname{Pr}\left\{\gamma_{e q} \leq \phi \gamma_E\right\} \nonumber\\
&=\int_0^{\infty} F_{{\gamma}_{\mathcal{D}}}\left(\phi \gamma\right) f_{\gamma_{\mathcal{E}_{\mathcal{Q}}}}\left(\gamma\right) d \gamma \nonumber\\ & \times  (1-F_{{\gamma}_{\mathcal{R}}}(\phi-1))+F_{{\gamma}_{\mathcal{R}}}(\phi-1).
\end{align}
Substituting Eqs. (\ref{cdffsov}) and (\ref{pdfev2}) into (\ref{eqn:soplower2}) and utilizing \cite[Eq. (2.24.1.1)]{brychkov1986integrals} yields
\begin{align}
\label{eqn:soplower22}
{SOP}^{II} &={\mathcal{S}_1} \sum_{m_h=1}^{\beta_h}\sum_{m_g=1}^{\beta_g}\sum_{{m_g}_e=1}^{{\beta_g}_e}{\mathcal{S}_2} {b_{m_h}^2} b_{m_g} b_{{m_g}_e}\nonumber\\&\times \mathrm{G}_{8r+1,8r+1}^{6r+1,6r}\left[{\mathcal{B}_1}\frac{{{\gamma_2}}}{{\gamma_{{e}_2}}\phi}\biggl| \begin{array}{c}
{1-l_{h_2},1-l_{g_2}, 1,l_{g_7},l_{h_1}} \\
l_{h_2}, l_{g_8}, 0,1-l_{h_1},1-l_{g_1}
\end{array}\right] \nonumber \\ &\times (1-F_{{\gamma}_{\mathcal{R}}}(\phi-1))+F_{{\gamma}_{\mathcal{R}}}(\phi-1),
\end{align}
where ${\mathcal{B}_1}=\left(\frac{B_{g_e}}{B_g}\right)^r$, ${\mathcal{S}_1}=\frac{\xi_h^4 \xi_g^2 {\xi_g}_e^2 {A_h}^2 {A_g} A_{{g}_e}}{2^{2r}}$, ${\mathcal{S}_2}=\frac{{r}^{2\alpha_h+\alpha_g+{\alpha_g}_e+2 m_h+m_g+{m_g}_e-5}}{{4\pi}^{4(r-1)}}$, $l_{g_7}=\left(\frac{1+{\xi_{g_e}}^2}{r},\ldots,\\ \frac{1+{\xi_{g_e}}^2+r-1}{r}\right)$, $l_{g_8}=\left(\frac{{\xi_{g_e}}^2}{r},  \ldots, \frac{{\xi_{g_e}}^2+r-1}{r}, \frac{\alpha_{g_e}}{r},\ldots, \frac{\alpha_{g_e}+r-1}{r},\frac{m_{g_e}}{r}, \ldots, \frac{m_{g_e}+r-1}{r}\right)$, and $F_{{\gamma}_{\mathcal{R}}}(\phi-1)=\sum_{n=0}^{\infty} \frac{(-1)^n\left(\frac{1}{b \sqrt{{\gamma}_1}}\right)^{a+n+1}({\phi-1})^\frac{a+n+1}{2}}{n !\left(a+n+1\right)}$.

\textbf{\textit{Scenario-III:}}
The lower bound of SOP for RIS-aided mixed RF-FSO framework with simultaneous eavesdropping attack via the RF and FSO links can be defined as 
\begin{align}
\label{ssop3}
{SOP}^{III}= 1-(SOP_1\times SOP_2),
\end{align}
where
\begin{align}
\label{eqn:soplower31} 
SOP_1= 1-\int_0^{\infty} F_{\gamma_{\mathcal{R}}}\left(\phi \gamma\right) f_{{\gamma}_{\mathcal{E}_{\mathcal{P}}}}\left(\gamma\right) d \gamma, \\ \label{eqn:soplower32} 
SOP_2= 1-\int_0^{\infty} F_{\gamma_{\mathcal{D}}}\left(\phi \gamma\right) f_{{\gamma}_{\mathcal{E}_{\mathcal{Q}}}}\left(\gamma\right) d \gamma.
\end{align}
By placing \eqref{eq2} and \eqref{pdfev1} into \eqref{eqn:soplower31} and utilizing \cite[Eq. (3.326.2)]{gradshteyn1988tables}, $SOP_1$ is expressed as
\begin{align}
SOP_1&=1-\frac{1}{b_e^{a_e+1}{\gamma_{e_{1}}}^{\frac{a_e+1}{2}}\Gamma(a_e+1)}\nonumber\\& \times \sum_{n=0}^{\infty} \frac{(-1)^n \left(\frac{\phi^\frac{1}{2}}{b \sqrt{{\gamma}_1}}\right)^{a+1+n}\left({b_e\sqrt{{\gamma_{{e}_1}}}}\right)^{T_1}\Gamma(T_1)}{n !\left(a+n+1\right)\Gamma(a+1)},
\end{align}
where $T_1=(a+a_e+n+2)$. Plugging (\ref{cdffsov}) and (\ref{pdfev2}) into \eqref{eqn:soplower32}, utilizing \cite[Eq. (2.24.1.1)]{brychkov1986integrals} to conduct integration and facilitating the expression, $SOP_2$ is obtained as
\begin{align}
SOP_2=&1-{\mathcal{S}_1} \sum_{m_h=1}^{\beta_h}\sum_{m_g=1}^{\beta_g}\sum_{{m_g}_e=1}^{{\beta_g}_e}{\mathcal{S}_2} {b_{m_h}^2} b_{m_g} b_{{m_g}_e}\nonumber\\\times&\mathrm{G}_{8r+1,8r+1}^{6r+1,6r}\left[{\mathcal{B}_1}\frac{{{\gamma_2}}}{{\gamma_{{e}_2}}\phi}\biggl| \begin{array}{c}
{1-l_{h_2},1-l_{g_2}, 1,l_{g_7},l_{h_1}} \\
l_{h_2}, l_{g_8}, 0,1-l_{h_1},1-l_{g_1}
\end{array}\right].
\end{align}

\subsection{Strictly Positive Secrecy Capacity Analysis}

The probability of SPSC serves to be one of the critical performance metrics that assures a continuous data stream only when the secrecy capacity remains a positive value in order to maintain secrecy in optical wireless communication. Mathematically, probability of SPSC is characterized as \cite[Eq.~(25)]{islam2020secrecy}
\begin{align}
\label{eqn:spsc_1}
SPSC&=\Pr\left\{C_{s}>0\right\}=1-SOP|_{R_{s}=0}.
\end{align}
Formulation of SPSC is readily obtained through substitution of the SOP formula from \eqref{eqn:sop3}, \eqref{eqn:soplower22}, and \eqref{ssop3} into \eqref{eqn:spsc_1}. Hence,
\begin{align}
\label{eqn:spsc_I}
{SPSC}^{I}&=1-{SOP}^{I}|_{R_{s}=0}, \text{(Scenario-I)}
\\
\label{eqn:spsc_II}
{SPSC}^{II}&=1-{SOP}^{II}|_{R_{s}=0}, \text{(Scenario-II)}
\\
\label{eqn:spsc_III}
{SPSC}^{III}&=1-{SOP}^{III}|_{R_{s}=0}. \text{(Scenario-III)}
\end{align}


\subsection{Effective Secrecy Throughput (EST)}

The EST is a gauge of performance indicator that clearly incorporates both tapping channel dependability and indemnity constraints. It essentially measures the mean rate at which secure data gets transmitted from the source to the destination with no interception. EST can be expressed numerically as
 \cite[Eq.~(5)]{lei2020secure}
\begin{align}
\label{eqn:EST}
EST=R_{S}(1-SOP).
\end{align}
Formulation of EST is readily obtained through substitution of the SOP formula in \eqref{eqn:EST}. Hence,
\begin{align}
{EST}^{I}&=R_{S}(1-{SOP}^{I}),   \text{ (Scenario-I)}
\\
{EST}^{II}&=R_{S}(1-{SOP}^{II}),  \text{ (Scenario-II)}
\\
{EST}^{III}&=R_{S}(1-{SOP}^{III}). \text{ (Scenario-III)}
\end{align}

\subsection{Intercept Probability (IP)}
IP refers to the likelihood that an eavesdropper successfully intercepts the data maintained at the actual receiving device. It represents the likelihood that secrecy capacity is greater than zero.
\\
\textbf{\textit{Scenario-I:}}
The IP for \textit{Scenario-I} is obtained as \cite[Eq.(23)]{sarker2021intercept}
\begin{align}
\label{IP_1}
{IP}^{I}=\int_0^{\infty} F_{{\gamma}_{\mathcal{R}}}(\gamma)f_{{\gamma}_{\mathcal{E}_{\mathcal{P}}}}(\gamma) d \gamma.
\end{align}
Plugging (\ref{eq2}) and (\ref{pdfev1}) into (\ref{IP_1}), performing integration utilizing \cite[Eq. (3.326.2)]{gradshteyn1988tables} and simplifying the expression, ${IP}^{I}$ is obtained as
\begin{align}
IP^{I}&=\frac{1}{b_e^{a_e+1}{\gamma_{e_{1}}}^{\frac{a_e+1}{2}}\Gamma(a_e+1)}\nonumber\\& \times \sum_{n=0}^{\infty} \frac{(-1)^n \left(\frac{1}{b \sqrt{{\gamma}_1}}\right)^{a+n+1}\left({b_e\sqrt{{\gamma_{{e}_1}}}}\right)^{T_1}\Gamma(T_1)}{n !\left(a+n+1\right)\Gamma(a+1)}.
\end{align}
\textbf{\textit{Scenario-II:}}
Similar to ${IP}^{I}$,  ${IP}^{II}$ can be expressed as
\begin{align}
\label{IP_2}
{IP}^{II}=\int_0^{\infty} F_{{\gamma}_{\mathcal{D}}}(\gamma)f_{{\gamma}_{\mathcal{E}_{\mathcal{Q}}}}(\gamma) d \gamma.
\end{align}
Plugging (\ref{cdffsov}) and (\ref{pdfev2}) into (\ref{IP_2}), performing integration utilizing \cite[Eq. (2.24.1.1)]{brychkov1986integrals} and simplifying the expression, ${IP}^{II}$ is obtained as
\begin{align}
{IP}^{II}&={\mathcal{S}_1} \sum_{m_h=1}^{\beta_h}\sum_{m_g=1}^{\beta_g}\sum_{{m_g}_e=1}^{{\beta_g}_e}{\mathcal{S}_2} {b_{m_h}^2} b_{m_g} b_{{m_g}_e}\nonumber\\&\times\mathrm{G}_{8r+1,8r+1}^{6r+1,6r}\left[{\mathcal{B}_1}\frac{{{\gamma_2}}}{{\gamma_{{e}_2}}}\biggl| \begin{array}{c}
{1-l_{h_2},1-l_{g_2}, 1,l_{g_7},l_{h_1}} \\
l_{h_2}, l_{g_8}, 0,1-l_{h_1},1-l_{g_1}
\end{array}\right].
\end{align}
\textbf{\textit{Scenario-III:}}
Using \eqref{eqn:spsc_III}, IP for simultaneous eavesdropping attacks via the RF and FSO links can be defined as
\begin{align}
\label{ip3}
{IP}^{III}= 1-{SPSC}^{III}.
\end{align}


\section{Numerical Results}

In this section, we present some numerical results based on the deduced analytical expressions of secrecy performance indicators i.e. SOP, SPSC, IP, ASC, and EST. Our analytical results are demonstrated for all three eavesdropping scenarios. To corroborate those analytical outcomes, we also exhibit MC simulations via generating Rician and Málaga random variables in MATLAB. This simulation is performed by averaging 100,000 channel realizations to get each secrecy indicator value. All analytical results are performed by considering specific parametric values such as $K\geq1$, $\Omega\geq1$, $\mathcal{N}\geq2$, $\gamma_{1}\geq0$, $\gamma_{2}\geq0$, $\gamma_{e_{1}}\geq0$, $\gamma_{e_{1}}\geq0$, $0\leq$ $ R_{s}$(bits/sec/Hz) $\leq1$, $(\alpha,\beta)$=($2.296,2$) for strong turbulence, ($4.2,3$) for moderate turbulence, and ($8,4$) for weak turbulence, $r=(1,2)$, and $\xi=(1.1,6.7)$.

The impact of average SNR on EST for a range of target secrecy rate i.e. $R_{s}$ is investigated in Figs. \ref{EST_1f} and \ref{EST_3f}.
\begin{figure}  [!ht]
\vspace{0.00mm}   \centerline{\includegraphics[width=0.35\textwidth,angle =0]{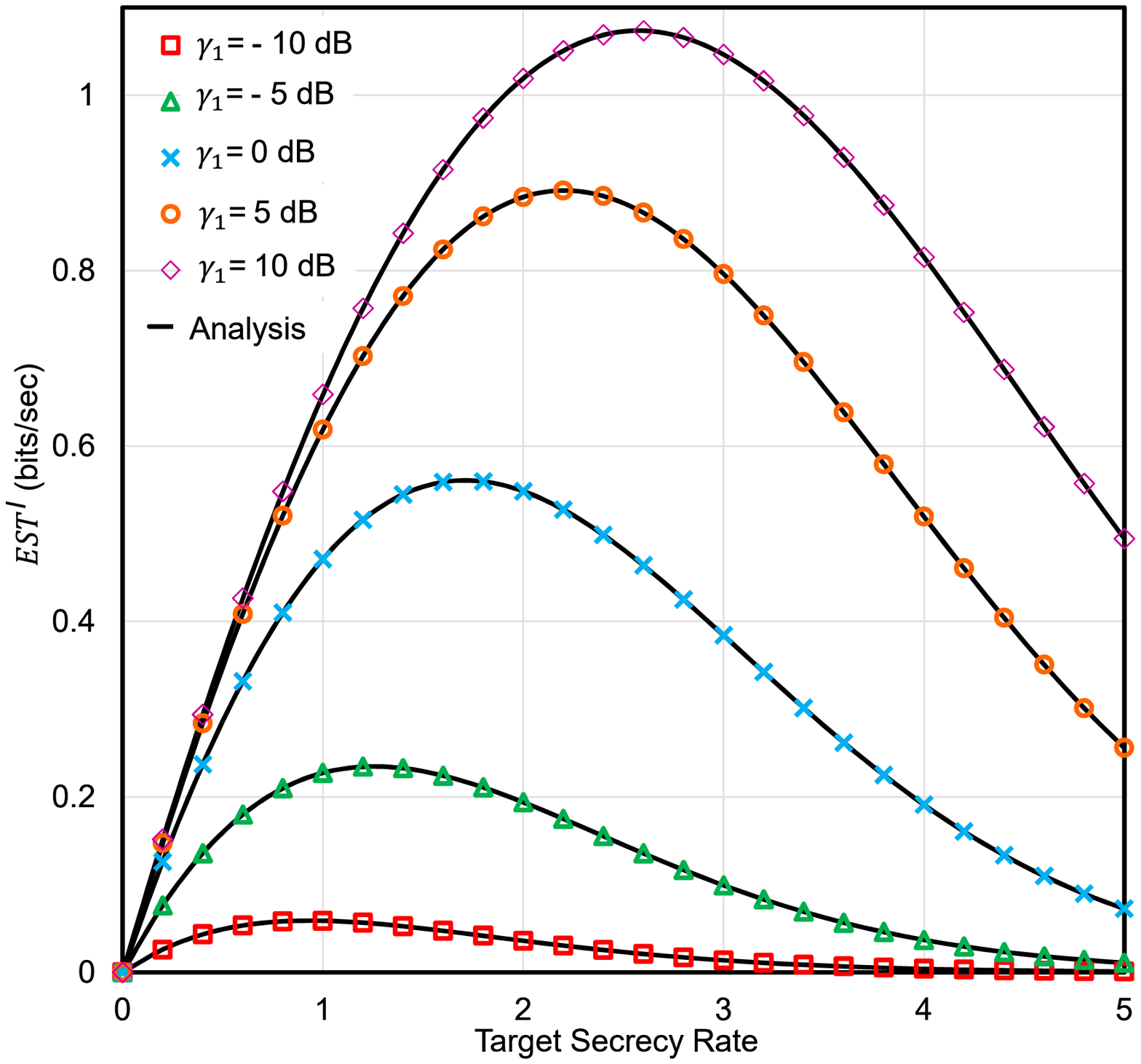}}
    \vspace{0mm}
    \caption{
         The ${EST}^{I}$ versus $R_{s}$ for selected values of $\gamma_{1}$ with $K_{1}=K_{2}=K_{3}=2$, $\Omega_{1}=\Omega_{2}=\Omega_{3}=3$, $\alpha_{h}=\alpha_{g}=2.296$, $\beta_{h}=\beta_{g}=2$, $\xi_{h}=\xi_{g}= 6.7$, $r=1$, $\mathcal{N}_1=2$, $\gamma_{2}=10$ dB, $\gamma_{{e}_1}=-5$ dB, and $R_s=0.1$.
    }
    \label{EST_1f}
\end{figure} 
\begin{figure}  [!ht]
\vspace{0.00mm}
    \centerline{\includegraphics[width=0.35\textwidth,angle =0]{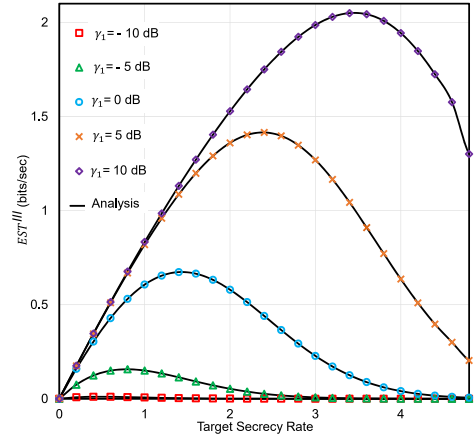}}
    \vspace{0mm}
    \caption{
         The ${EST}^{III}$ versus $R_{s}$ for selected values of $\gamma_{1}$ with $K_{1}=K_{2}=K_{3}=2$, $\Omega_{1}=\Omega_{2}=\Omega_{3}=2$, $\alpha_{h}=\alpha_{g}=2.296$, $\beta_{h}=\beta_{g}=2$, $\xi_{h}=\xi_{g}= 1.1$, $r=1$, $\mathcal{N}_1=5$, $\gamma_{2}=10$ dB, $\gamma_{{e}_1}=-5$ dB, and $R_s=0.1$.
    }
    \label{EST_3f}
\end{figure} 
Both results are formulated by varying the average SNR of the RF main channel i.e. $\gamma_{1}$, from $-10$ dB to $10$ dB, considering Fig. \ref{EST_1f} represents \textit{Scenario-I} (${EST}^{I}$) and Fig. \ref{EST_3f} stands for \textit{Scenario-III} (${EST}^{III}$). It is observed that both figures devise concave down-shaped curves where EST rises to a certain value of $R_{s}$ then declines afterward. Several reasons are responsible for this peculiar event. As the lower value of $R_{s}$ needs lesser security maintenance resources that results in higher EST for the system, higher $R_{s}$ creates the inverse outcome. Furthermore, larger $R_{s}$ exacerbates the channel condition by introducing additional noise, interference, and fading in the system that has a notable impact on EST. Supposedly, the EST vs $R_{s}$ relation in Figs. \ref{EST_1f} and \ref{EST_3f} demonstrate the optimization between the secured transmission rate of the system and the required resources for maintaining security.

The impression of shape parameter ($K$) and scale parameter ($\Omega$) of Rician fading distribution on the secrecy performance of the proposed channel is studied in Figs. \ref{IP_1f} and \ref{IP_3f}.
\begin{figure}  [!ht]
\vspace{0.00mm}
    \centerline{\includegraphics[width=0.35\textwidth,angle =0]{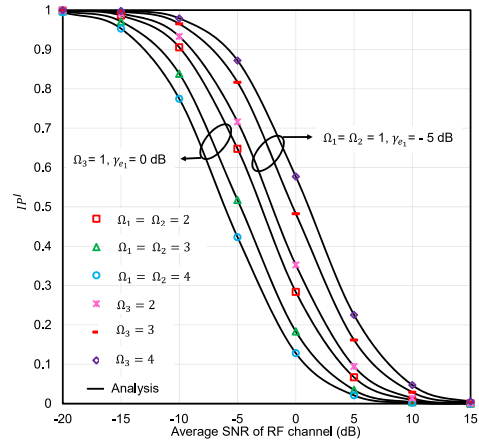}}
        \vspace{0mm}
    \caption{
          The ${IP}^{I}$ versus $\gamma_{1}$ for selected values of $\Omega_{1}, \Omega_{2}, \Omega_{3}$, and $\gamma_{{e}_1}$ with $K_{1}=K_{2}=K_{3}=2$, $r=1$, $\mathcal{N}_1=2$, and $R_s=0.1$.}
    \label{IP_1f}
\end{figure} 
\begin{figure}  [!ht]
\vspace{0.00mm}
    \centerline{\includegraphics[width=0.35\textwidth,angle =0]{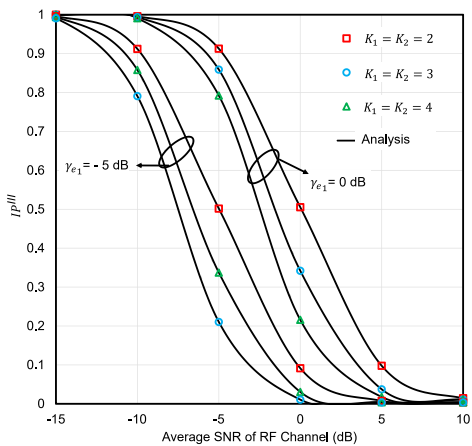}}
        \vspace{0mm}
    \caption{
          The ${IP}^{III}$ versus $\gamma_{1}$ for selected values of $K_{1}, K_{2}$, and $\gamma_{{e}_1} $ with $\alpha_{h}=\alpha_{g}=\alpha_{{g}_e}=2.296$, $\beta_{h}=\beta_{g}=\beta_{{g}_e}=2$, $\xi_{h}=\xi_{g}=\xi_{{g}_e}=6.7$, $K_{3}=2$, $\Omega_{1}=\Omega_{2}=\Omega_{3}=3$, $r=1$, $\mathcal{N}_1=4$, $\gamma_{2}=25$ dB, $\gamma_{{e}_1}=0$ dB, and $R_s=0$.}
    \label{IP_3f}
\end{figure} 
In Fig. \ref{IP_1f}, ${IP}^{I}$ in \textit{Scenario-I} decreases when $\Omega_{1}$ and $\Omega_{2}$ from $\mathcal{S}-\mathcal{L_{P}}-\mathcal{R}$ link is boosted from $2$ to $4$; therefore, secrecy performance increases. This circumstance occurs because a higher scale parameter value improves signal quality by reducing signal attenuation, and so makes the communication channel more reliable. For the same reason, secrecy performance downturns when $\Omega_{3}$ rises from $2$ to $4$ as it declines security of the main channel by strengthening $\mathcal{S}-\mathcal{L_{P}}-\mathcal{E_{P}}$ link. Additionally, the channel with larger shape parameter value experiences a lesser fading effect because the line-of-sight signal is much stronger than the scattered signal when $K$ is high. This phenomenon is justified in Fig. \ref{IP_3f} for \textit{Scenario-III} where system performance improves for the higher values of $K_{1}$ and $K_{2}$. Moreover, both figures support that better quality of channel gained by increasing Rician fading parameters is more significant when the average SNR of $\mathcal{S}-\mathcal{L_{P}}-\mathcal{E_{P}}$ link i.e. $\gamma_{{e}_1}$ is lower.

Fig. \ref{SOP_1f} studies a comparison between two detection techniques i.e. HD and IM/DD for \textit{Scenario-I}.
\begin{figure}  [!ht]
\vspace{0.00mm}
\centerline{\includegraphics[width=0.35\textwidth,angle=0]{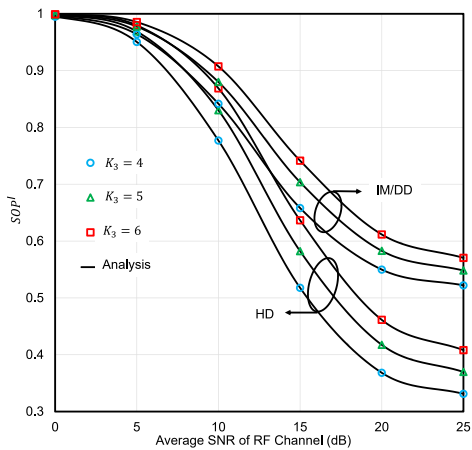}}
        \vspace{0mm}
    \caption{
         The ${SOP}^{I}$ versus $\gamma_{1}$ for selected values of $K_{3}$ and r with $\Omega^{\prime}=1$, $K_{1}=K_{2}=2$, $\Omega_{1}=\Omega_{2}=\Omega_{3}=3$, $\alpha_{h}=\alpha_{g}=2.296$, $\beta_{h}=\beta_{g}=2$, $\xi_{h}=\xi_{g}= 6.7$, $\mathcal{N}_1=2$, $\gamma_{2}=25$ dB, $\gamma_{{e}_1}=10$ dB, and $R_s=0.1$.}
    \label{SOP_1f}
\end{figure}
Conventionally, HD technique has the ability of frequency shifting in a high frequency range. Thus, HD is less susceptible to wiretapping and contain more secrecy advantages than IM/DD for secured wireless channel. Fig. \ref{SOP_1f} upholds this agreement and our analysis supports the results demonstrated in \cite{islam2020secrecy}. It can also be observed from Fig. \ref{SOP_1f} that higher value of $K_{3}$ lessens the fading effect of $\mathcal{S}-\mathcal{L_{P}}-\mathcal{E_{P}}$ link; hence makes the wiretapping capability stronger while reducing the system performance.

The impact of FSO eavesdropper for \textit{Scenario-II} is represented in both Figs. \ref{EST_2f} and \ref{SOP_2f}.
\begin{figure}  [!ht]
\vspace{0.00mm} \centerline{\includegraphics[width=0.35\textwidth,angle =0]{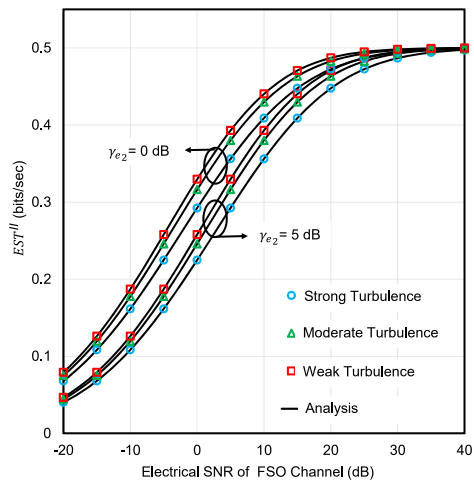}}
        \vspace{0mm}
    \caption{
         The ${EST}^{II}$ versus $\gamma_{2}$ for selected values of $\alpha_{h}, \alpha_{g}, \beta_{h}, \beta_{g},$ and $\gamma_{{e}_2}$ with $\alpha_{{g}_e}= \beta_{{g}_e}=1 $, $\xi_{h}=\xi_{g}={\xi_g}_e=1.1$, $r=1$, $K_{1}=K_{2}=2$, $\mathcal{N}_1= 4$, $\Omega_{1}=\Omega_{2}=3$, $\gamma_{1}=10$ dB, and $R_s=0.5$.
         }
    \label{EST_2f}
\end{figure}
\begin{figure}  [!ht]
\vspace{0.00mm}
    \centerline{\includegraphics[width=0.35\textwidth,angle=0]{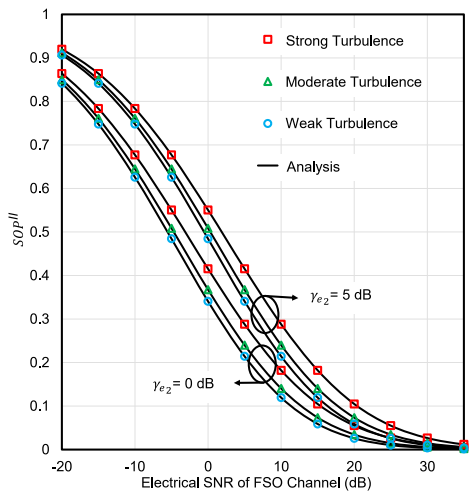}}
        \vspace{0mm}
    \caption{
         The ${SOP}^{II}$ versus $\gamma_{2}$ for selected values of $\alpha_{h}, \alpha_{g}, \beta_{h}, \beta_{g}$, and $\gamma_{{e}_2}$ with $\alpha_{{g}_e}= \beta_{{g}_e}=1 $, $\xi_{h}=\xi_{g}={\xi_g}_e=1.1$, $r=1$, $K_{1}=K_{2}=2$, $\mathcal{N}_1=4$, $\Omega_{1}=\Omega_{2}=3$, $\gamma_{1}=10$ dB, and $R_s=0.5$.}
    \label{SOP_2f}
\end{figure} 
It is observed that $EST^{II}$ in Fig. \ref{EST_2f} is higher for $\gamma_{e_{2}}=0$ dB while comparing with $\gamma_{e_{2}}=5$ dB. For the same case, $SOP^{II}$ in Fig. \ref{SOP_2f} is lower for $\gamma_{e_{2}}=0$ dB and slightly better for $\gamma_{e_{2}}=5$ dB. These results are obvious as higher SNR of $\mathcal{R}-\mathcal{L_{Q}}-\mathcal{E_{Q}}$ link for Fig. \ref{pm1} (\textit{Scenario-II}) constantly increases the wiretapping capability of $\mathcal{E_{Q}}$ that ultimately reduces the secrecy performance.

\begin{figure}  [!ht]
\vspace{0.00mm}
    \centerline{\includegraphics[width=0.35\textwidth,angle =0]{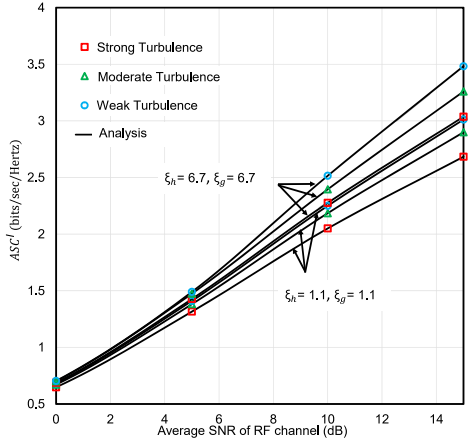}}
        \vspace{0mm}
    \caption{
          The ${ASC}^{I}$ versus $\gamma_{1}$ for selected values of $\alpha_{h}, \alpha_{g},\beta_{h},\beta_{g}, \xi_{h}$, and $\xi_{g}$ with $K_{1}=K_{2}=K_{3}=1$, $\Omega_{1}=\Omega_{2}=\Omega_{3}=2$, $r=1$, $\mathcal{N}_1=2$, $\gamma_{2}=25$ dB, $\gamma_{{e}_1}=-5$ dB, and $R_s=0.1$.}
    \label{ASC_1f}
\end{figure} 
The influence of the pointing error of M\'alaga distribution on the secrecy performance is studied in Figs. \ref{ASC_1f}-\ref{SOP_3f} for all three scenarios in Fig. \ref{pm1}. Results clearly demonstrate that secrecy performance for three eavesdropping scenarios upturns when the pointing error index i.e. $\xi$ increases from $1.1$ (severe pointing error state) to $6.7$ (negligible pointing error state). This outcome is consistent while varying both the average SNRs of the RF link (Figs. \ref{ASC_1f} and \ref{SOP_3f}) and FSO link (Fig. \ref{IP_2f}). It is a notable fact that ${SOP}^{III}$ in Fig. \ref{SOP_3f} downturns when $\mathcal{N}_{1}$ jumps from $1$ to $4$. This incident proves that the presence of a higher number of reflecting elements for the RIS-state fading channel gradually increases the system performance.

The effect of atmospheric turbulence conditions on EST, SOP, IP (\textit{Scenario-II}), and ASC (\textit{Scenario-I}) is investigated in Figs. \ref{EST_2f}-\ref{IP_2f}. It can be observed that regardless of the different wiretapping scenarios, the secrecy performance of the proposed system in Fig. \ref{pm1} improves when the natural turbulence condition of the M\'alaga link shifts from strong turbulence to weak turbulence (i.e., the lower SOP, IP or the higher EST, ASC).

\begin{figure}  [!ht]
\vspace{0.00mm}
    \centerline{\includegraphics[width=0.35\textwidth,angle =0]{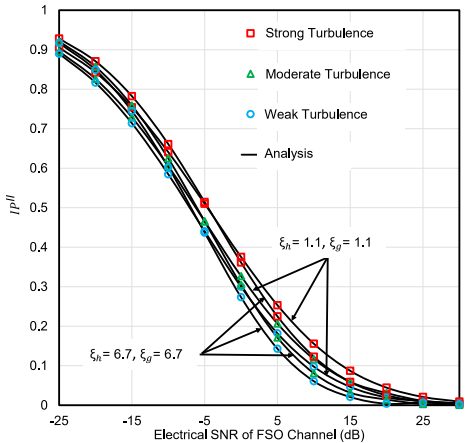}}
        \vspace{0mm}
    \caption{
         The ${IP}^{II}$ versus $\gamma_{2}$ for selected values of $\alpha_{h}, \alpha_{g}, \beta_{h}, \beta_{g}, \xi_{h}$, and $\xi_{g}$ with $\alpha_{{g}_e}= \beta_{{g}_e}=1 $, $K_1=K_2=2$, $\Omega_1=\Omega_2=2$, $r=1$, $\mathcal{N}_1=4$, and $R_s=0.5$.
         }
\label{IP_2f}
\end{figure} 
\begin{figure}  [!ht]
\vspace{0mm}
    \centerline{\includegraphics[width=0.35\textwidth,angle =0]{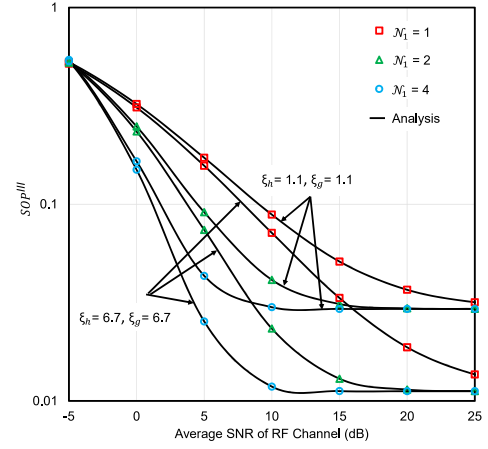}}
        \vspace{0mm}
    \caption{
         The ${SOP}^{III}$ versus $\gamma_{1}$  for selected values of $ \mathcal{N}_1, \xi_{h}$ and $\xi_{g}$ with $\alpha_{h}=\alpha_{g}=\alpha_{{g}_e}=2.296$, $\beta_{h}=\beta_{g}=\beta_{{g}_e}=2$, $\xi_{{g}_e}=6.7$, $r=1$, $K_{1}=K_{2}=K_{3}=1$, $\Omega_{1}=\Omega_{2}=\Omega_{3}=2$, $\gamma_{2}=20$ dB, $\gamma_{{e}_1}=-5$ dB, and $R_s=0.1$.
         }
    \label{SOP_3f}
\end{figure} 

A comparative analysis of \textit{Scenario-I}, \textit{Scenario-II} and \textit{Scenatio-III} for the proposed system model is presented in Fig. \ref{SPSC_c}.
\begin{figure}  [!ht]
\vspace{0.00mm}
    \centerline{\includegraphics[width=0.35\textwidth,angle=0]{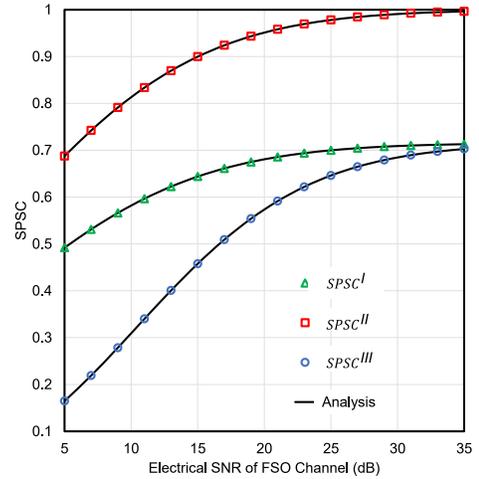}}
        \vspace{0mm}
    \caption{
         The ${SPSC}$ versus $\gamma_{2}$ for different scenarios with $\alpha_{h}=\alpha_{g}=2.296$, $\beta_{h}=\beta_{g}=2$, $\alpha_{{g}_e}= \beta_{{g}_e}=1$, $\gamma_{1}=2$ dB, $\gamma_{{e}_1}=\gamma_{{e}_2}=-1$ dB, ,$\xi_{h}=\xi_{g}={\xi_g}_e={\xi_h}_e=1.1$, $r=1$, $K_{1}=K_{2}=K_{3}=2$, $\mathcal{N}_1=3$, and $\Omega_{1}=\Omega_{2}=\Omega_{3}=2$.}
    \label{SPSC_c}
\end{figure} 
Here SPSC is plotted against the average SNR of $\mathcal{R}-\mathcal{L_Q}-\mathcal{D}$ link i.e. $\gamma_2$ using the expressions denoted in \eqref{eqn:spsc_I}-\eqref{eqn:spsc_III}. Conventionally, an FSO link is more secure and has lower susceptibility to eavesdropping compared to the RF link \cite{sarker2021effects}. This statement is validated in our analysis in Fig. \ref{SPSC_c} where ${SPSC}^{II}$ demonstrates the highest probability of SPSC among the three scenarios. Likewise, \textit{Scenario-I} is worse than \textit{Scenario-II} as an RF link is usually more vulnerable to wiretapping than an FSO link. However, the worst case is displayed by ${SPSC}^{III}$ (\textit{Scenario-III}) in Fig. \ref{SPSC_c}. The reason behind this case is both $\mathcal{E_P}$ and $\mathcal{E_Q}$ remain active simultaneously in \textit{Scenario-III} that eventually creates the strongest form of eavesdropping for the channel in Fig. \ref{pm1}.

\section{Conclusions}

This study aimed at analyzing the security performance of a DF-based RIS-aided RF-FSO communication system in the presence of wiretapping attacks in both RF and FSO networks. We derived closed-form expressions for various performance metrics, such as ASC, SPSC, EST, IP, and lower-bound SOP, to efficiently evaluate the impact of each parameter on the secrecy performance. The study validated its analytical outcomes using MC simulations. Numerical results reveal that fading severity, pointing errors, and natural turbulence parameters have a significant impact on secrecy performance. The study also explores the trade-off between the target secrecy rate and the resources required to maintain security via EST. Moreover, the result highlights the superiority of HD over IM/DD for optical signal detection. We also conducted a comparative analysis of three proposed wiretapping scenarios and concluded that simultaneous wiretapping has a more detrimental effect on the secrecy performance than individual wiretapping. Finally, it is claimed that the FSO link is less susceptible to wiretapping than the RF link.

\bibliographystyle{IEEEtran}
\bibliography{IEEEabrv,main.bib}

\end{document}